\documentclass[aps,superscriptaddress,nobibnotes,reprint]{revtex4-1}

\usepackage{amsmath,amssymb,dsfont,relsize}
\usepackage{graphicx}% Include figure files
\usepackage{dcolumn}% Align table columns on decimal point
\usepackage{bm}% bold math
\usepackage{latexsym,tikz,hyperref,url}
\usepackage{array}
\usepackage{setspace}
\usepackage{xcolor}
\usepackage{amsthm}

\DeclareMathAlphabet{\mathpzc}{OT1}{pzc}{m}{it}

\makeatletter
\renewcommand{\p@subsection}{}
\renewcommand{\p@subsubsection}{}
\makeatother

\begin{document}

\title{Geometric characterization of SARS-CoV-2 pandemic events} 
\author{Ivan~Bonamassa}
\email{ivan.bms.2011@gmail.com}
\affiliation{Department of Physics, Bar-Ilan University, 52900 Ramat-Gan, Israel}
\author{Marcello Calvanese Strinati}
\affiliation{Department of Physics, Bar-Ilan University, 52900 Ramat-Gan, Israel}
\author{Adrian Chan}
\affiliation{Department of Physics, Bar-Ilan University, 52900 Ramat-Gan, Israel}
\author{Ouriel~Gotesdyner}
\affiliation{Department of Physics, Bar-Ilan University, 52900 Ramat-Gan, Israel}
\author{Bnaya~Gross}
\affiliation{Department of Physics, Bar-Ilan University, 52900 Ramat-Gan, Israel}
\author{Shlomo~Havlin}
\affiliation{Department of Physics, Bar-Ilan University, 52900 Ramat-Gan, Israel}
\author{Mario~Leo}
\affiliation{Dipartimento di Matematica e Fisica ``Ennio de Giorgi'', Universit\`a del Salento, Lecce, Italy}
\date{\today}

\begin{abstract}
While the SARS-CoV-2 keeps spreading world-wide, comparing its evolution across different nations is a timely challenge of both theoretical and practical importance. 
The large variety of dissimilar and country-dependent epidemiological factors, in fact, makes extremely difficult to understand their influence on the epidemic trends within a unique and coherent framework. 
We present a geometric framework to characterize, in an integrated and low-dimensional fashion, the epidemic plume-like trajectories traced by the infection rate, $I$, and the fatality rate, $D$, in the $(I,D)$ plane. 
Our analysis enables the definition of an %data-driven 
epidemiometric system based on three geometric observables rating the SARS-CoV-2 pandemic events via scales analogous to those for the magnitude and the intensity of seismic events. 
Being exquisitely geometric,
%Due to its exquisitely geometric nature, 
our framework can be applied to classify other epidemic data and secondary waves, raising the possibility of designing epidemic alerts or early warning systems to enhance public and governmental responses to a rapidly emerging outbreak. 
%We expect that our results will help enhancing public and governmental responses to curb epidemic events since their early stages by means of epidemic alerts and warning systems, further providing with unfamiliar perspectives the design of models for longer-termed forecasting of real-world outbreaks. 
\end{abstract}

\maketitle

%\section{Introduction} \label{sec:1}
% Like hurricanes, earthquakes or tornados, pandemics represent another class of catastrophic events with the potential of disrupting the delicate equilibrium of our modern society. 
%\emph{\underline{Introduction}.} 
%Since its identification in late 2019, the SARS-CoV-2 pandemic has caused more than half a million loss in human lives and brought nations into severe economic crises. %disclosing the tremendous damage that a pandemic can have on the globalized world~\cite{}. 
The unprecedented amount of epidemic data collected worldwide on SARS-CoV-2 raises nowadays a unique opportunity to quantify, in a way analogous to other extreme events~\cite{rodriguez2007handbook,blaikie2014risk}, the catastrophic impact that a pandemic can have on the globalized world~\cite{farrell2020will,raza2020coronavirus}. 
In the context of earthquakes, for example, the existence of the Richter~\cite{richter1935instrumental} and Mercalli~\cite{mercalli1902sulle} measures, quantifying respectively the magnitude and the intensity of a local seismic event, has helped policy-makers to take informed decisions yielding better intervention strategies (e.g.\ by means of tsunami alerts or rapid post-earthquakes notifications~\cite{kanamori1997real,aranda1995mexico,kanamori2005real,horiuchi2005automatic}) and strong governmental actions (e.g.\ investments in anti-seismic infrastructures~\cite{reitherman2012earthquakes,achaoui2016seismic}) to prevent their potential impact. 
Similarly, in meteorology, the Fujita~\cite{fujita1971proposed} and the Saffir-Simpson~\cite{saffir1973hurricane,simpson1974hurricane} scales have offered researchers with heuristic measures to estimate the potential damage inflicted by, respectively, tornados and hurricanes on human-build structures and vegetation, raising the opportunity of designing ever-increasingly refined early warning systems and alert protocols~\cite{barber2001tornado,sorensen2000hazard,montz2017natural}. 
In the realm of pandemics, however, metric systems enabling a comprehensive classification of their types have (to the best of our knowledge) never been proposed, resulting in a fundamental gap in the human fight against this type of catastrophic events. \\
%Despite the large volume of research on epidemic spreading~\cite{} and the tremendous efforts invested in the modeling and forecasting the SARS-CoV-2 evolution~\cite{}, a metric system for systematically quantifying the strength of pandemic types has (to the best of our knowledge) never been proposed, resulting into a fundamental gap in the human fight against this type of catastrophic events.
\indent
The theoretical and practical implications of this important and timely challenge are numerous. 
Disposing of a robust and comprehensive framework to classify %the {\em force} and the {\em intensity} of 
the SARS-CoV-2 pandemic events reported across different countries not only can enhance early~\cite{pluchino2020novel,kogan2020early} public and governmental responses in containing the spreading and/or better absorbing the impact of a rapidly emerging epidemic outbreak, but it can further provide new information to better understand real-world epidemics and to boost the forecasting power of existing models~\cite{perc2020forecasting,zlatic2020bi,petropoulos2020forecasting,fanelli2020analysis,sornette2020interpreting,pollett2020identification,dehning2020inferring,vespignani2020modelling,cintra2020mathematical}.\\
\indent 
A fundamental difficulty to achieve this goal relies in the large heterogeneity of epidemiological and country-dependent factors characterizing the global pandemic trends. 
%At the time of writing, in fact, the nations' epidemic trends exhibit a strong variety of spreading patterns, with a few of them already ahead of the pandemic outbreak, like Switzerland, Austria or New Zealand~\cite{} others experiencing resurgent secondary waves, like Iran, Sweden or Israel~\cite{qi2020model}, and other ones instead still on their ways towards the infection peaks.
Diverse isolation~\cite{maier2020effective} and social distancing strategies~\cite{ferguson2006strategies,germann2006mitigation,adolph2020pandemic,morse2006next,mate2020evaluating,meidan2020alternating,karin2020adaptive,block2020social,aleta2020evaluation}, age, gender impact~\cite{alon2020impact,wenham2020covid} and demographic characteristics of different populations~\cite{goldstein2020demographic}, local transportation systems~\cite{gross2020spatio,chinazzi2020effect,kraemer2020effect,aleta2020data,schlosser2020covid}, tracking and testing policies~\cite{moscovitch2020better,segal2020building}, health systems' capacities~\cite{grasselli2020critical} and many other factors, make difficult the design of quantitative epidemiometric systems for country-to-country comparison~\cite{roda2020difficult,loeffler2020covid,ognyanova2020state}. 
Moreover, epidemic models or inference algorithms fine-tuned to this constellation of features, inevitably result into theoretical or semi-empirical frameworks whose complexity rapidly increases with the large number of data-driven parameters considered~\cite{squillante2020attacking,castro2020predictability,arenas2020mathematical,bertozzi2020challenges,petropoulos2020forecasting,singer2020short,ribeiro2020short,orea2020effective,mora2020semiempirical}.\\ 
%In these mathematical and/or computational tour-de-forces, low-dimensional parametrizations grasping only the essentials of the complex dynamics observed in real-world data would be extremely helpful and largely simplify the analysis.\\
\indent 
In this work, we present a geometric, low-dimensional method to classify the impact the SARS-CoV-2 pandemic events observed across different nations. 
After performing a statistical best-fit of the epidemic data for the infected, $I$, and deceased, $D$, rates, we analyze the geometry of their plume-like trajectories in the $(I,D)$ plane. 
Moving to a polar representation, we classify the plumes' form through a set of {\em three} geometric parameters yielding two complementary rating scales for the SARS-CoV-2 pandemic types: one according to their {\em epidemic magnitude}---labeled with roman numbers from $\mathds{I}$ to $\mathds{X}$ for increasing strengths---and measuring the ``size'' of a national outbreak, and a second one according to their {\em intensity}---labeled alphabetically from $\mathds{A}$ to $\mathds{D}$ for increasing speed---quantifying instead the damage inflicted on the population. 
Even though each country exhibits its own pandemic fingerprint, our geometric method unveils hidden similarities shared by their global trends emerging from an integrated representation of their evolution. 
We further provide a qualitative understanding of the epidemiological information contained in the developed epidemic measures, and discuss the theoretical and practical implications of our results.
%to performing a similar analysis on the trajectories of a well-mixed SEIRD model~\cite{}.\\ 
%For instance, we find that, although US has experienced a type V pandemics, its damage over the population is of class B, similar to countries like Greece or Finland. \\
%This allows to define a scale of pandemic intensities that largely enriches the spectrum of real-world pandemic types beyond the two simplistic notions of ``slow'' and ``fast'' epidemics. \\
%Although our approach is still rudimentary, we only a preliminary step towards more refined classification of pandemic types. We believe our findings will inspire further research the design of better alerts and intervention strategies prevention, and that it will further help reaching optimized models for real-world epidemics. 
%In a Bayesian inference scenario~\cite{}, for example, learning the data from a lower-dimensional parametrization may help finding better priors for more efficient and longer-term forecasting.\\ 

%\section{Comparing the pandemic trends} \label{sec:2}
\emph{1. \underline{Comparing the pandemic trends}.} 
To set the stage for a cross-comparison of the countries' pandemic trends, let us consider two widely reported epidemiological observables, i.e.\ the infected ($I$) and the fatality ($D$) daily rates. 
These quantities can vary strongly from one country to another, depending on a wide variety of factors. 
Even within the same country, the numbers may fluctuate from one day to another due to delays in reporting, transitions to new surveillance and/or tasting systems, or simply because of weekly periodic variations in the number of daily tests performed. 
Smoothening the data under a suitable moving average window unveils the trends of the time series, enabling a preliminary comparison. \\
\begin{figure}[b]\vspace*{-0.5cm}
	\includegraphics[width=\linewidth]{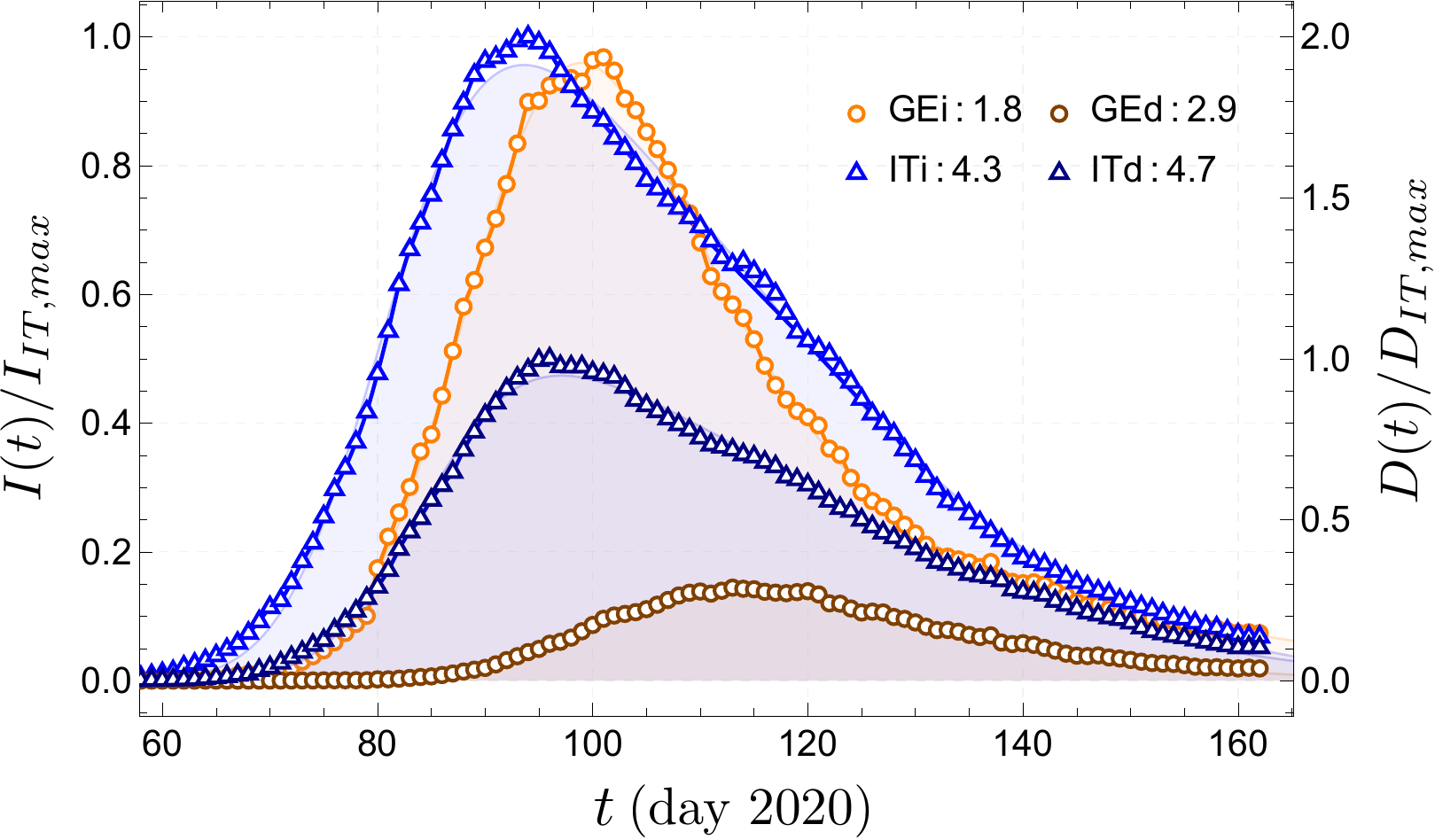}\vspace*{-0.2cm}
	\caption{\textbf{Temporal evolution of epidemic profiles.} (Color online) 
	Comparison between SARS-CoV-2 infected and fatality rate profiles for Italy (respectively, blue and dark blue triangles) and Germany (respectively, orange and dark orange circles) under a $15$-days moving average.
	Infected and deceased rates have been normalized with respect to the corresponding largest values reported in Italy. Different $y$ scales have been chosen for clarity of exposition. 
	(Legend) Skewnesses values obtained after best fitting the global trends via skewness Gaussian distributions. Data taken from Ref.~\cite{repo}.}
	\label{fig:skewness}	
\end{figure}
\indent 
As a demonstrative example, let us consider the pandemic trends reported in Italy and in Germany (Fig.~\ref{fig:skewness}). 
In both countries, it took approximately $6$ weeks for the outbreaks to reach their infection peaks, with similar fast-rising trends and comparable numbers of newly infected patients per day. 
Their post-peak behaviors, however, differ noticeably: whereas Germany's infection curve has decayed almost as quickly as it rose, resulting in a reduction by $50\%$ of the daily infected in nearly $15$ days, it has taken almost double this time to Italy to reach similar conditions. 
This is nicely reflected in the values of the skewnesses calculated after best-fitting the data with asymmetric Gaussians (Fig.~\ref{fig:skewness}, legend), showing a decay of the Italian trend roughly $2.3$ times slower than the one observed in Germany. 
Slow infection rate decays similar to those reported in Italy have also been observed in the United States, the United Kingdom or Russia and can be explained as the result of new regional outbreaks spreading throughout the country after the national lockdown, hinting at a difficulty in identifying/containing the virus since its early stages. 
%Taken in isolation, nevertheless, the numbers of infected do not reveal the whole story. 
Additional information can be found by performing a similar analysis of the fatality rates. 
This quickly reveals that Italy had to face much more critical conditions, counting (on average) at least $3$ times more fatalities per day than Germany, with a death peak located only $4$ days after the infected one (by contrast with the $2$ weeks lag reported in Germany) and a skewness roughly twice the one measured based on the German data (Fig.~\ref{fig:skewness}, legend). 
\begin{figure}[t]
	\includegraphics[width=0.95\linewidth]{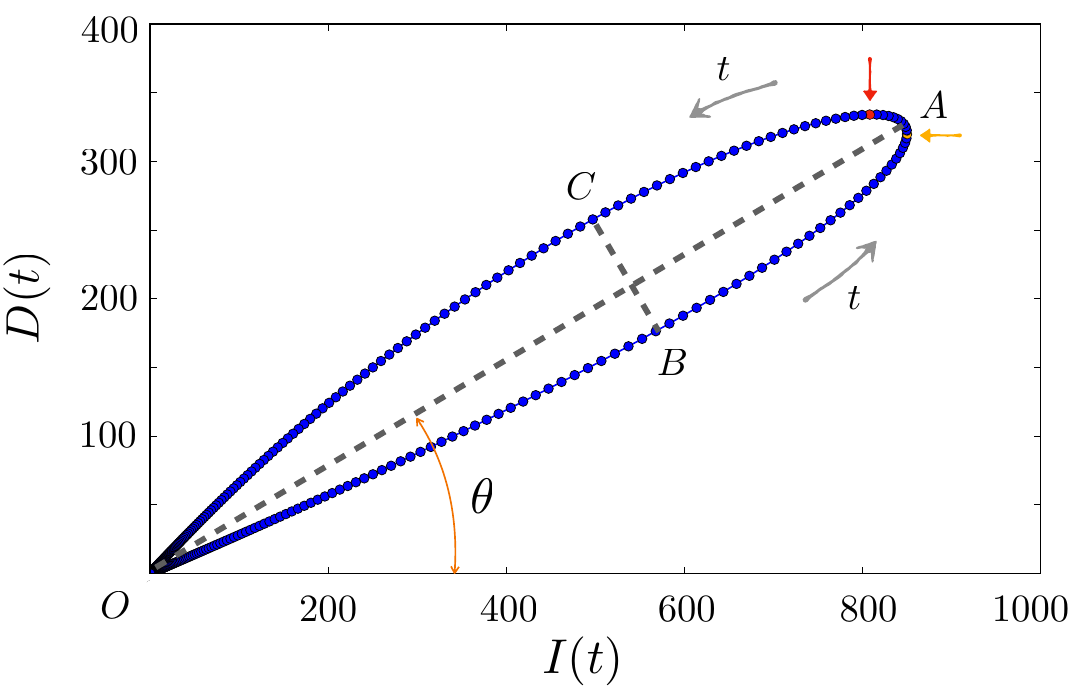}\vspace*{-0.25cm}
	\caption{\textbf{Geometric parametrization of typical epidemic trends.} (Color online) Demonstration of our geometric method to analyze a typical epidemic trajectory in the $(I,D)$ plane; in this representation, time flows counter-clockwise. 
	Yellow and red arrows point respectively at the infected and fatality rate peaks, while the markers evolve on a daily rate. 
	We calculate the largest radius, $r_{max}\equiv \overline{AO}$, its inclination, $\theta$, and the largest width, $r_\perp\equiv\overline{BC}$, of the plume-like trajectory traced by the epidemic evolution.
	These quantities define the triple $(r_{max},\theta,\rho)$, with $\rho\equiv r_\perp/r_{max}$, used to analyze  classify events as described in the text.}
	\label{fig:demo}	\vspace*{-0.5cm}
\end{figure}
\noindent 
By looking at these differences separately, one can heuristically conclude that both countries have experienced outbreaks with similar magnitudes though causing very different damages on the population, with Germany applying a more efficient policy of containment and/or testing and with Italy reaching very critical level in its health system. \\
\indent 
To extend the comparison to other countries, it proves essential to identify a suitable metric system grasping and systematically quantifying the relevant information (e.g.~peak values, infected-fatality peaks lags, post-peak decay rates, etc.) enshrined in the evolution profiles of the $I$ and $D$ trajectories. 
In the spirit of dynamical systems theory~\cite{strogatz2018nonlinear}, we approach this problem by departing from the representation of the dynamical observables $I(t)$ and $D(t)$ as functions of time, focusing instead on their mutual evolution in the $(I,D)$ plane (Fig.~\ref{fig:demo}). 
In this way, a comprehensive portrait of the outbreak dynamics of each country can be described by an integrated plume-like trajectory (Fig.~\ref{fig:demo}) whose geometric features, as we shall see in what below, provide an exhaustive and low-dimensional classification of the SARS-CoV-2 pandemic types according to their magnitude and intensity.

%\iffalse
%\begin{figure}
%	\centering
%	\begin{tikzpicture}[      
%	every node/.style={anchor=north east,inner sep=0pt},
%	x=1mm, y=1mm,
%	]   
%	\node (fig1) at (0,0)
%	{\includegraphics[scale=0.34]{skewness_err_new}};
%	\end{tikzpicture}
%	\caption{\textbf{Skewness Parameters.} The skewness parameters accounting for the  daily infected and deceased rates. To get better statistics, the skewness of each country has been averaged over the range of 2-20 days window sizes. The standard deviation is also shown. Most of the countries perform similarly and located at the center of the phase diagram but some outliers appear.}
%	\label{fig:skewness_phase}	
%\end{figure}

\begin{figure*}
\centering
	\hspace*{-0.5cm}
	\includegraphics[width=0.9\linewidth]{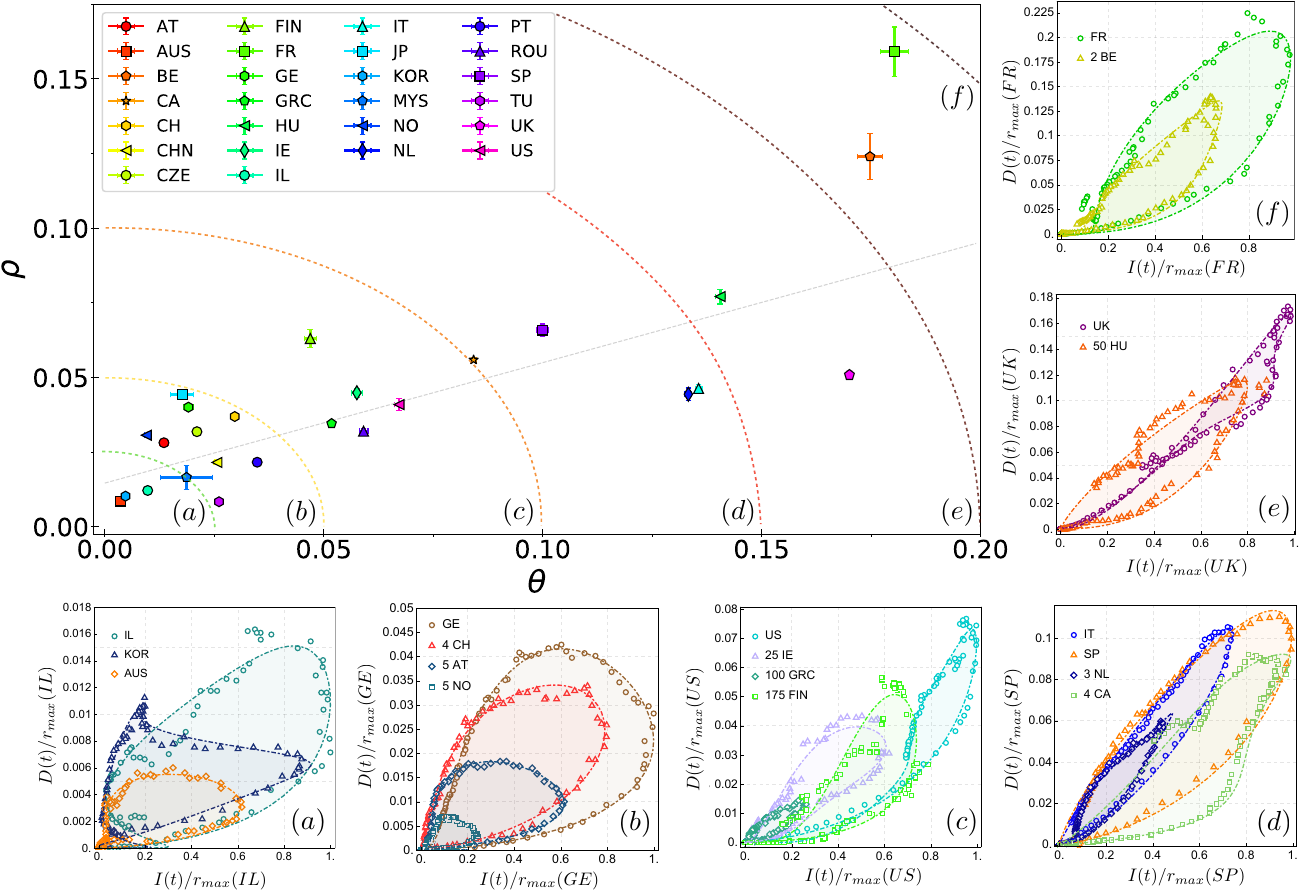}\vspace*{-0.25cm}
	\caption{\textbf{$(\theta,\rho)$-parametrization of typical SARS-CoV-2 pandemic trajectories.} (Color online)
	\textbf{a})--\textbf{f}) Epidemic trajectories (dot-dashed curves) in the $(I,D)$ plane obtained by interpolating the fitted epidemic states of each country describing their first-wave events. 
	These are generated via the distribution functions best fitting the infected and fatality rates (as in Fig.~\ref{fig:skewness}) ruling the evolution of the real-world epidemic states (markers). 
	The $(\theta,\rho)$ phase diagram summarizes the results of our geometric analysis, where error bars are calculated by considering a confidence interval of $99\%$ over values of $\theta$ and $\rho$ for trajectories fitting epidemic data under moving $n$-day averages with $n$ from $1$ to $15$. 
	A linear regression analysis yields $\rho\simeq \rho_0+\alpha \theta$, with intercept $\rho_0=(1.5\pm0.5)\times10^{-2}$ and slope $\alpha=0.40\pm0.06$ (dashed gray line). 
	The linear dependence of $\rho$ on $\theta$ is supported by a Pearson coefficient $\tilde{r}\simeq0.81$~\cite{cox1979theoretical}. 
	Countries in the scatter plot are then selected according to shells (dashed circles) of increasing radiuses, identifying six heuristic regimes, ranging from \textbf{a}) the ideal scenario of slow outbreaks with good tracking intervention and low fatality rates, to \textbf{f}) extreme cases representing very fast outbreaks with high case-fatality rates and critical conditions of their health systems. 
	\textbf{a})--\textbf{f}) Sampled trajectories belonging to each heuristic group exhibit epidemic angles (notice the different extend of the $y$-axes) whose similarities can be better highlighted by normalizing the epidemic data of each group via the largest $r_{max}$ of the country in the collection. 
	The normalization factors are respectively: \textbf{a}) $r_{max}(IL)\simeq5.2\times10^2$, \textbf{b}) $r_{max}(GE)\simeq5\times10^3$, \textbf{c}) $r_{max}(US)\simeq 3\times10^4$, \textbf{d}) $r_{max}(SP)\simeq7\times10^3$, \textbf{e}) $r_{max}(UK)\simeq5.3\times10^3$, \textbf{f}) $r_{max}(FR)\sim4.2\times10^3$.} 
	\label{fig:theta_rho}	\vspace*{-0.2cm}
\end{figure*} 

\begin{figure}[t]
	\centering
	\includegraphics[width=0.85\linewidth]{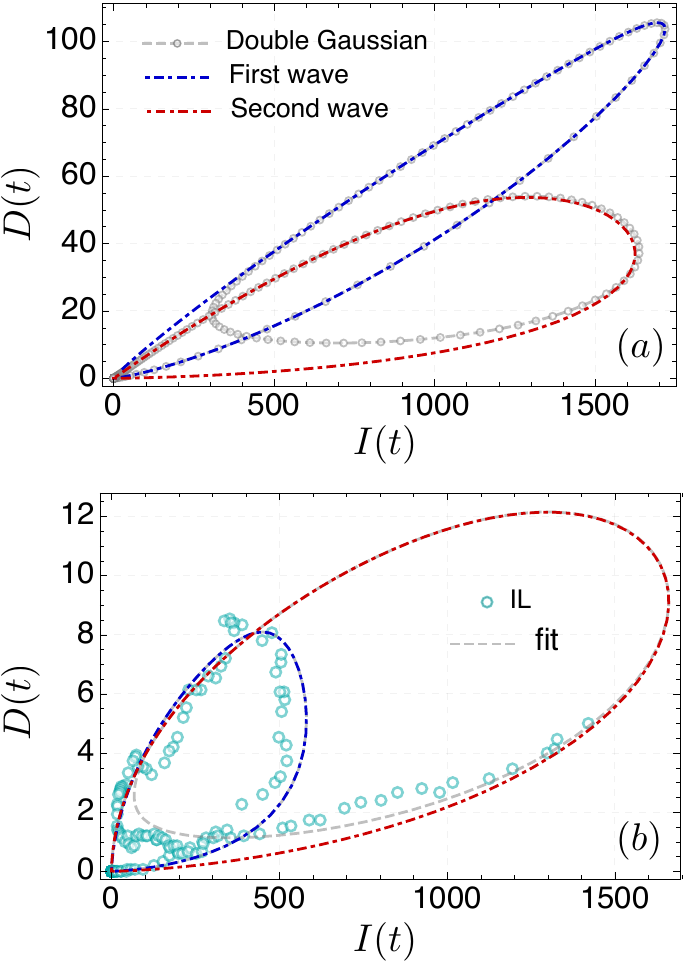}\vspace*{-0.3cm}
	\caption{\textbf{Second-wave analysis.} 
	\textbf{a}) Synthetic data (gray circles) generated by considering for both the infected and the fatality rates bimodal skewed Gaussians with well resolved first- and second-wave peaks.
	The two trajectories, describing respectively the first (blue dot-dashed curve) and second (red dot-dashed curve) waves, are analyzed by separating the corresponding fitting plumes. 
	Both the epidemic events have same magnitude %($\mathds{III}$ if $P\sim10^8$, $\mathds{VI}$ if $P\sim10^7$ or larger magnitude for smaller population sizes) 
	but different epidemic angles, with the first wave of category $\mathds{B}$ and the second one of category $\mathds{A}$. 
	\textbf{b}) Epidemic data describing the evolution of the outbreak in Israel (darker cyan circles) and its fitted double-plume trajectory (gray dashed curve). 
	By separating each fitting curve as in \textbf{a}) we identify the first epidemic wave (blue, dot-dashed curve) of magnitude $\mathpzc{T}_1\simeq4.9$ and category $\mathds{A}^{\boldsymbol{-}}$, and the ongoing second wave (red, dot-dashed curve). 
	As of July 20 2020, the second-wave event in Israel has epidemic angle $\theta_{IL}\simeq0.005$ (i.e.\ category $\mathds{A}$), and magnitude $\mathpzc{T}_2\simeq6.6$ (i.e.\ class $\mathds{VII}$). 
	}\label{fig:secwav}\vspace*{-0.5cm}
\end{figure}

\emph{2. \underline{Geometric parametrization}.} 
In the $(I,D)$ plane, the daily epidemic state of each country traces a trajectory that, after departing from the healthy phase $(0,0)$, leaves behind its own epidemic fingerprint.
For reasons that will be clarified below, let us call ``typical'' an epidemic trajectory analogous to the one depicted in Fig.~\ref{fig:demo} and characterized by a dynamics that unfolds counterclockwise in the plane, reaching first the infected peak, then the fatality peak, and finally heads back towards the axes origin. 
Notice that during this heading back regime, new outbreaks could emerge due to e.g.\ premature lifting of the lockdown measures, pushing the epidemic state to trace a different plume-like trajectory (Fig.~\ref{fig:secwav}\textbf{a}) describing a new epidemic event. 
Here, we focus our analysis on the epidemic data reported during the first-wave events in countries that have passed both their infected and fatality rate peaks. 
In the discussion section we will see how our geometric method can be naturally extended to analyze second-wave events like those observed in United States, Iran or Israel (see Fig.~\ref{fig:secwav}\textbf{b}).\\ 
\indent 
To quantitatively compare the countries' typical epidemic trajectories, we introduce three geometric parameters, $(r_{max},\theta,\rho)$, that we measure after transforming the epidemic observables $I,\,D$ into polar coordinates. 
We define with $r_{max}$ the maximal radius of the epidemic trajectory, with $\theta$ the angle formed by $r_{max}$ with respect to the $I$-axes, and finally with $\rho=r_{\perp}/r_{max}$ the relative width of the plume-like curve, where $r_{\perp}$ is its maximal width (Fig.~\ref{fig:demo}). 
While $r_{max}$ measures the largest extent of the epidemic plume---and has therefore units of population---the quantities $\theta$ and $\rho$ have a genuine geometric nature and disclose different information about the {\em intensity} of an epidemic event. 
Large values of $\theta$, in fact, reflect a fast raise of the number of fatalities jointly with a rapid increase of the newly infected, which would occur in cases of a critical health system but also reflect the country's demographic features, like age and morbidity distributions, social interactions, etc. 
In this respect, we adopt $\theta$ as an estimator of the epidemic ``speed'', with large (small) angles describing fast (slow) outbreaks.
The parameter $\rho$, on the other hand, can be written in terms of the ``eccentricity'', $e$, of the epidemic trajectory as $\rho=\sqrt{1-e^2}$~\cite{nota1}, so that
%increasing values of $\rho$ describe plumes with decreasing time lags between the infected and fatality rates peaks. 
decreasing values of $\rho$ characterize narrower plumes.
%This situation would naturally reflect late identification of infected or seriously ill patients---peak-to-peak proximity hints in fact at the existence of correlations among the reported infected and fatalities---resulting in an overflow of a country's medical system. 
This would naturally reflect situations of rapid patient identification, resulting in lower critical conditions of the country's hospitals and hence to less fatalities, i.e.\ lower values of the angle $\theta$. 
%This further hints at some degree of correlation between $\theta$ and $\rho$, with high values of $\rho$ describing situations of unsuccessful lockdown and/or tracking strategies, causing in turn a more critical health system and hence more fatalities, i.e.\ larger values of $\theta$. 
The results in Fig.~\ref{fig:theta_rho} support this heuristic interpretation, disclosing a linear relation between the two geometric observables $\theta$ and $\rho$. 
In light of the latter, we have selected countries according to their Euclidean distance from the origin of the $(\theta, \rho)$ scatter plot, yielding a preliminary partitioning of their outbreaks according to their speed, as shown in Fig.~\ref{fig:theta_rho}~\textbf{a})--\textbf{f}). 
Besides highlighting similar inclination of the epidemic trajectories, the normalization by $r_{max}$ adopted in Fig.~\ref{fig:theta_rho}~\textbf{a})--\textbf{f}) further discloses an additional degree of similarity between countries based on the time lag separating infected and fatality rate peaks. 
Countries with low angles like e.g.\ Germany, Austria or Norway (Fig.~\ref{fig:theta_rho}\textbf{b}) feature, in fact, round plumes with well separated peaks as well as more narrow forms like those reported in e.g.\ Israel (Fig.~\ref{fig:theta_rho}\textbf{a}), Greece or Finland (Fig.~\ref{fig:theta_rho}\textbf{c}). 
On the contrary, countries with large epidemic angles always correspond to narrow plumes with strong peak-to-peak proximity, as observed e.g.\ in Italy (Fig.~\ref{fig:theta_rho}\textbf{d}), Hungary (Fig.~\ref{fig:theta_rho}\textbf{e}) or Belgium (Fig.~\ref{fig:theta_rho}\textbf{f}). 
\begin{figure}[b]\vspace*{-0.25cm}
	\includegraphics[width=0.9\linewidth]{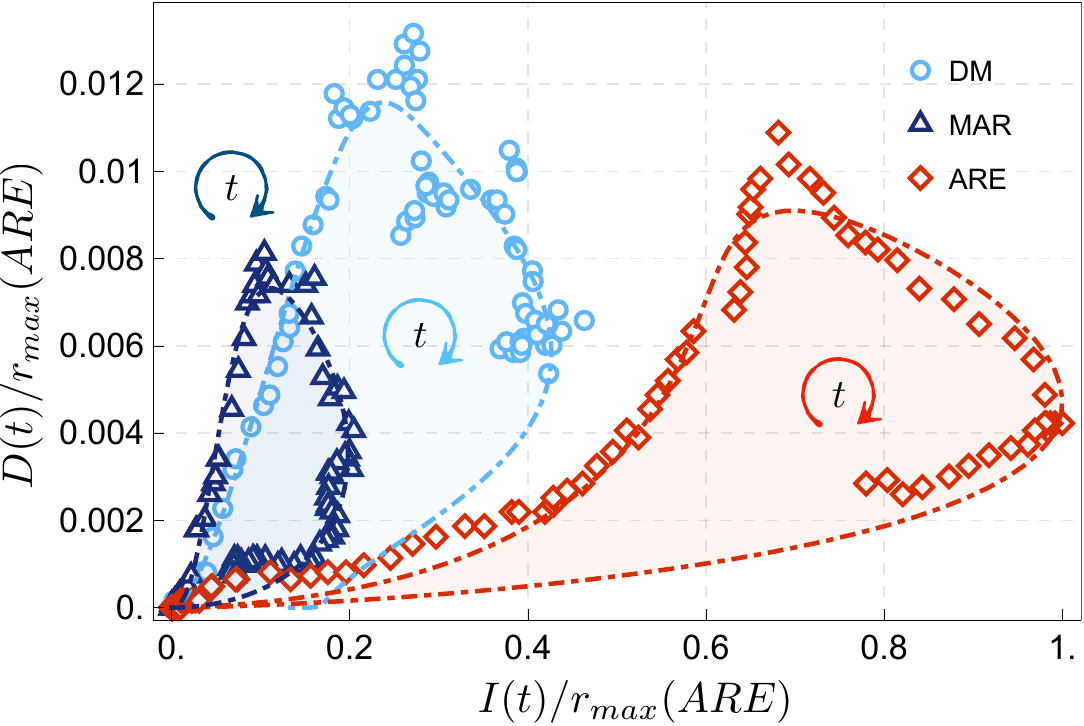}\vspace*{-0.2cm}
	\caption{\textbf{Non-typical SARS-CoV-2 trajectories.} (Color online) 
	Epidemic trajectories featuring clockwise evolution, due to a fatality peak preceding the infected one, as reported in Morocco, United Arab Emirates (ARE), Iran, Dominican Republic and other countries. 
	Here, $r_{max}(ARE)\simeq3.2\times10^2$.}
	\label{fig:anom}	
\end{figure}
\noindent 
%However, the normalized trajectories in the inset panels to Fig.~\ref{fig:theta_rho} depicts the opposite scenario, with small $\rho$ resulting into rounder and rounder trajectories (like those for Israel, Germany, Austria, or Norway in Figs.~\ref{fig:theta_rho}\textbf{a},\ \textbf{b}),while larger values of $\rho$ describing narrow pseudo-ellipses (reported in Figs.~\ref{fig:theta_rho}\textbf{d}--\textbf{f} for Italy, France, Spain and others). 
%This is nothing but a side effect of $r_{max}\gg r_\perp$ and a more meaningful comparison of the trajectories forms can be obtained by analyzing the eccentricities of their normalized traces. 
As we shall see in the next section, the eccentricity of the normalized epidemic plumes nicely grasps this important information, enabling to define a simple yet informative metric system characterizing the intensity of the SARS-CoV-2 types.\\
\indent
Before delving into the details of this classification, let us complete the picture by considering those epidemic plumes which have not been included in the analysis due to their {\em non}-{\em typical} evolution in the $(I,D)$ plane. 
This includes epidemic data describing first-wave events whose fatality rate peak has preceded the infected one, as observed e.g.\ in Brazil, Mexico, United Arab Emirates, Iran (Fig.~\ref{fig:anom}) and few others. 
Such peak inversion translates into a clockwise evolution of the epidemic trajectory in the $(I,D)$ plane, whose geometric features could, in principle, be analyzed according to our developed method but should not be similarly interpreted. 
Fatality-rate peaks preceding the infected ones can be, in fact, only explained as the result of sparse and incomplete reporting due to e.g.\ limited resources at early stages of diagnostics or absence of post-mortem identification. % or political moves to try obfuscating~\cite{theguardian2020global} the epidemic trends. 
For the sake of simplicity, in what follows we will focus only on the analysis of {\em typical} epidemic trajectories, while the classification of non-typical cases will be performed elsewhere.\vspace*{+0.02cm} 

\par
\emph{3. \underline{Classification of SARS-CoV-2 types}.} 
Having introduced a geometric parametrization of the countries typical epidemic trajectories, let us now focus on defining a suitable and conventional metric system to systematically quantify their magnitude and intensity.\vspace*{+0.02cm}

\begin{figure}[t]
	\includegraphics[width=0.92\linewidth]{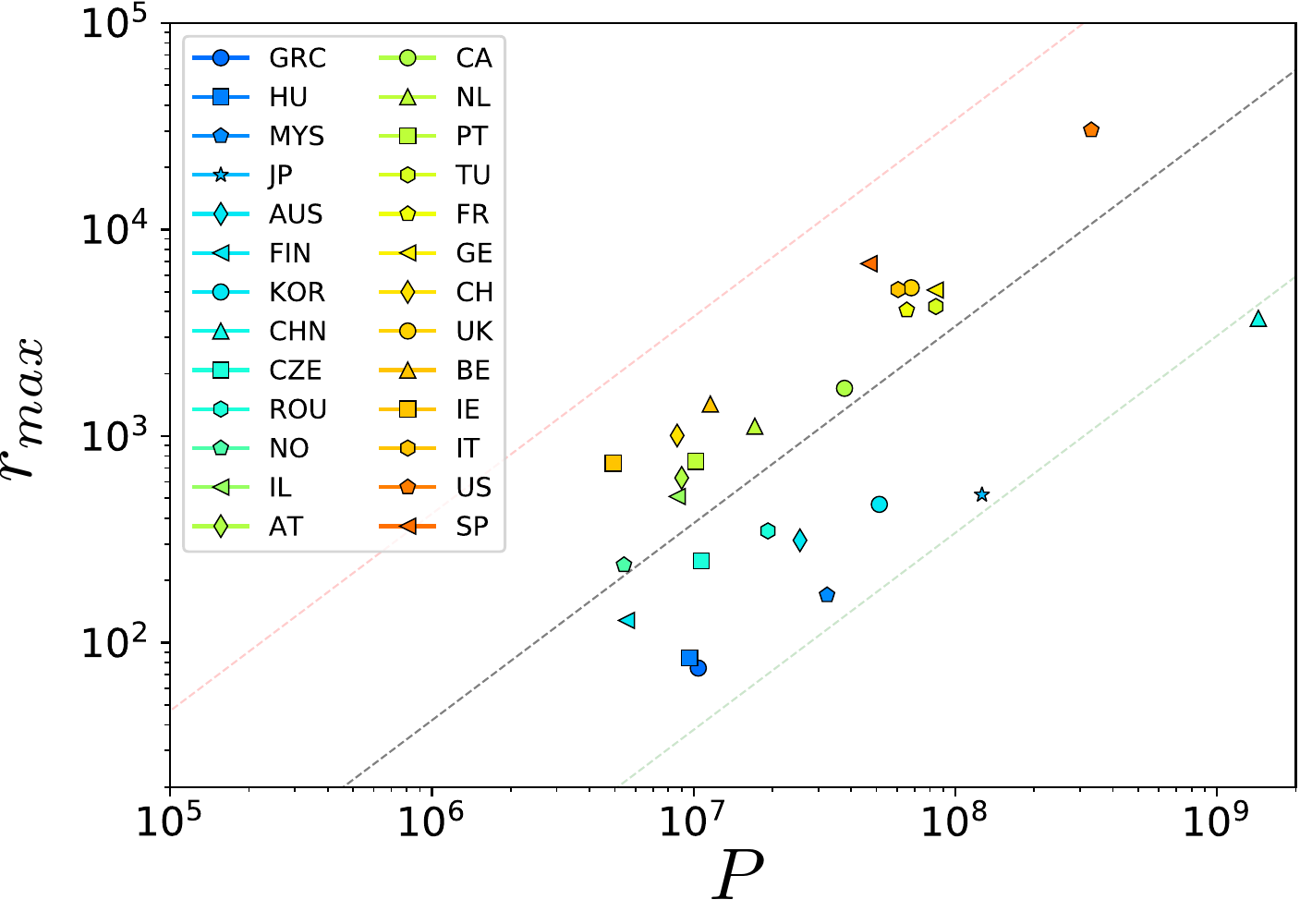}\vspace*{-0.2cm}
	\caption{\textbf{Extent/population correlation in SARS-CoV-2 trajectories.} (Color online) Distribution of $r_{max}$ vs.\ population size, $P$, for the countries listed in Fig.~\ref{fig:theta_rho}.
	A linear fit ($p$-value $\sim10^{-6}$) yields $\log(r_{max})=r_0+\beta \log(P)$, with intercept $r_0=-4.1\pm1.5$ and slope $\beta=0.95\pm0.20$ (dashed gray line). 
	%Notice that $r_0\simeq-4.1$ implies a mean percentage $\mathpzc{A}_0=10^{r_0}\simeq10^{-4}$ for the available data, i.e.\ a mean $0.001\%$ of the countries' populations has been daily reported infected and/or deceased at their largest extent.
	The statistical significance of the linear relation is further supported by a Pearson correlation coefficient $\tilde{r}\sim0.71$. 
	Data points are are colored from blue to red for increasing values of the epidemic magnitude $\mathpzc{T}$ (see Tab.~\ref{tab:mag}). 
	Dashed green and red lines respectively describe the boundary of the $\boldsymbol{++}$ and $\boldsymbol{--}$ categories, rating the efficacy of each country intervention strategies (we refer to the main text for details).}\vspace*{-0.5cm}
	\label{fig:corr}	
\end{figure}

\indent 
\emph{Outbreak magnitude.} As we anticipated, the parameter $r_{max}$ yields an integrated measure in the $(I,D)$ plane of the largest extent of an epidemic trajectory in units of population, offering the opportunity to analyze the dependence of the epidemic extent on the population size, $P$, of the country where it spread. 
By plotting the distribution of $r_{max}$ as a function of $P$ (Fig.~\ref{fig:corr}), we find an approximate power law $r_{max}= \mathcal{A}P^\beta$, with $\mathcal{A}\in(0,1)$ a country-dependent proportionality factor and $\beta$ an exponent close to one. 
The nearly linear relation can be explained by interpreting $\mathcal{A}$ as the largest fraction of daily infected and deceased reported for a given population size, suggesting a rudimentary yet informative scale to meaningfully quantify the epidemic magnitude of each country.
%This enables a quantitative classification of the pandemic relative magnitudes together with a rating system of the social-distancing and lockdown strategies adopted by each government. \\
%\indent 
To this aim, let us introduce the dimensionless parameter $\mathpzc{x}=\log r_{max}/\log P$. 
Because $r_{max}$ and $P$ are linearly proportional, this relation reads as 
\begin{equation}\label{eq:mag}
\mathpzc{x}=1+\log\mathcal{A}/\log P,
\end{equation}
where $P>1$, so that $\mathpzc{x}\in(-\infty,1)$ is a monotonically increasing function of the proportionality factor $\mathcal{A}\in(0,1)$. 
This fraction is bounded by two extreme cases: $\mathcal{A}\to0$ describing an extremely weak (nearly absent) outbreak with only few infected/deceased daily cases, and $\mathcal{A}\to1$ representing instead the unlikely event of a nearly full population infected/deceased in a single day. 
Similar to the Richter metric system for local seismic events, an epidemic magnitude scale can be conventionally defined by choosing a suitable, monotonically increasing function of $\mathpzc{x}$ in Eq.~\eqref{eq:mag}, where $\mathpzc{x}$ can be interpreted as the ``epidemic force'' of an outbreak, measuring the largest fraction of new infected {\em and} deceased reported in a single day. 
%Based on the available data, we suitably 
Let us therefore define the {\em epidemic magnitude} as 
\begin{equation}\label{eq:magT}
\mathpzc{T}(\mathpzc{x})=10^{\mathpzc{x}/\mathpzc{x}^*}, 
\end{equation}
\noindent 
where $\mathpzc{x}^*\equiv1-3/\log P$ plays the role of a characteristic epidemic strength. 
The choice of $\mathpzc{x}^*$ allows to assign the scale $\mathpzc{T}=10$ to a pandemic event with nearly $0.1\%$ of the population reported infected and/or deceased in a single day (i.e.\ $\mathcal{A}^*=10^{-3}$), an outcome of catastrophic proportions. 
%Based on the available data, for which $\mathpzc{x}\in[0.27,0.53]$, we suitably define the epidemic magnitude as $\mathpzc{T}=e^{10\mathpzc{x}}/20$, where the exponential part highlights the repartition between different classes while the normalization yields values of $\mathpzc{T}\in(0,10)$. 
In light of Eqs.~\eqref{eq:mag}--\eqref{eq:magT}, we introduce an epidemiometric system classifying epidemic events by increasing magnitudes that is described as follows:
\begin{itemize}
\item[$\mathds{I}$)] $\mathpzc{T}\in [0.0,0.9]$, {\em micro} events: very weak outbreaks having strengths $\mathpzc{x}<0$ or, equivalently, $r_{max}<1$, i.e.\ an average of less than $1$ infected case per day;
\item[$\mathds{II}$)] $\mathpzc{T}\in[1.0,1.9]$, {\em minor} events: weak outbreaks with non-negative strengths corresponding to values of $r_{max}\in[1,4)$ for small countries with $P\sim10^5$, and $r_{max}\in[1,32)$ for large countries with $P\sim10^8$; 
\item[$\mathds{III})$] $\mathpzc{T}\in[2.0,2.9]$, {\em light} events: epidemics featuring e.g.\ values of $r_{max}\in[4,9)$ if $P\sim10^5$, and $r_{max}\in[32,250)$ for large countries with $P\sim10^8$; 
\item[$\mathds{IV})$] $\mathpzc{T}\in[3.0,3.9]$, {\em mild} events: outbreaks characterized by e.g.\ $r_{max}\in[9,16)$ if $P\sim10^5$, and $r_{max}\in[250,10^3)$ if $P\sim10^8$; 
\item[$\mathds{V})$] $\mathpzc{T}\in[4.0,4.9]$, {\em moderate} events: epidemics with e.g.\ $r_{max}\in[16,25)$ for $P\sim10^5$, and $r_{max}\in[10^3,3\times3.1\times 10^3)$ for $P\sim10^8$; 
\item[$\mathds{VI})$] $\mathpzc{T}\in[5.0,5.9]$, {\em strong} events: epidemics with e.g.\ $r_{max}\in[25,36)$ for $P\sim10^5$ (i.e.\ averaged daily percentage of infected/deceased reaching peaks between $0.025\%$ to $0.036\%$ of the total population), and $r_{max}\in[3.1\times10^3,7.8\times10^3]$ for $P\sim10^8$ (peaks between $0.003\%$ and $0.008\%$ of $P$ per day);
\item[$\mathds{VII})$] $\mathpzc{T}\in[6.0,6.9]$, {\em very strong} events: epidemics characterized by values of $r_{max}\in[36,49)$ for small countries with $P\sim10^5$, and $r_{max}\in[7.8\times10^3,1.7\times10^4)$ for large countries with $P\sim10^8$; 
\item[$\mathds{VIII})$] $\mathpzc{T}\in[7.0,7.9]$, {\em violent} events: outbreaks featuring e.g.\ values of $r_{max}\in[49,64)$ for $P\sim10^5$, and $r_{max}\in[1.7\times10^4,3.3\times10^4]$ for $P\sim10^8$; 
\item[$\mathds{IX})$] $\mathpzc{T}\geq8.0$, {\em extreme} epidemic events featuring e.g.\ values of $r_{max}\geq 64$ for $P\sim10^5$, and $r_{max}>3.3\times10^4$ for $P\sim10^8$, i.e.\ outbreaks whose daily percentages of infected/deceased reach peaks respectively larger than $0.06\%$ and $0.03\%$ of the total population.
\end{itemize}

\begin{table}[b]\vspace*{-0.4cm}
\begin{tabular}{c||cc|c}
\centering 
country & $\mathpzc{T}$ & $d$ & type\\
\hline
GRC  & 2.9 & 0.7 & $\mathds{III}^{++}$  \\
HU & 3.0 & 0.6 & $\mathds{IV}^{++}$ \\
MYS & 3.1  & 0.8 & $\mathds{IV}^{++}$ \\
JP & 3.4 & 0.9 & $\mathds{IV}^{++}$ \\
AUS & 3.7 & 0.5 & $\mathds{IV}^{++}$ \\
FIN  & 3.7 & 0.2 & $\mathds{IV}^{+}$ \\
KOR  & 3.7 & 0.6 & $\mathds{IV}^{++}$ \\
CHN & 3.8 & 1.1 & $\mathds{IV}^{+++}$ \\
CZE  & 3.9 & 0.2 & $\mathds{IV}^{+}$ \\
ROU & 3.9 & 0.3 & $\mathds{IV}^{+}$ \\
NO & 4.3 & -0.1 & $\mathds{V}^{-}$ \\
IL  & 4.9 & -0.2 & $\mathds{V}^{-}$ \\
AT & 5.1 & -0.26 & $\mathds{VI}^{-}$
\end{tabular}
\begin{tabular}{c||cc|c}
\centering 
country & $\mathpzc{T}$ & $d$ & type\\
\hline
CA  & 5.1 & -0.1 & $\mathds{VI}^{-}$ \\
NL & 5.2 & -0.25 & $\mathds{VI}^{-}$  \\
PT & 5.2 & -0.29 & $\mathds{VI}^{-}$  \\
TU  & 5.4 & -0.17 & $\mathds{VI}^{-}$  \\
FR & 5.6 & -0.26 & $\mathds{VI}^{-}$  \\
GE  & 5.7 & -0.25 & $\mathds{VI}^{-}$  \\
CH & 5.8 & -0.49 & $\mathds{VI}^{-}$  \\
UK & 5.9 & -0.35 & $\mathds{VI}^{-}$  \\
BE  & 6.0 & -0.52 & $\mathds{VII}^{--}$  \\
IE  & 6.0 & -0.58 & $\mathds{VII}^{--}$  \\
IT  & 6.0 & -0.39 & $\mathds{VII}^{-}$  \\
US & 6.5 & -0.45 & $\mathds{VII}^{-}$ \\
SP  & 6.6 & -0.62 & $\mathds{VII}^{--}$  
\end{tabular}
\caption{\textbf{Magnitude of SARS-CoV-2 outbreaks.} Classification of the SARS-CoV-2 first-wave epidemic events according to their magnitude, $\mathpzc{T}=10^{\mathpzc{x}/\mathpzc{x}^*}$, where $\mathpzc{x}^*\equiv1-3/\log P$ represents an upper limit to the epidemic ``strength'' characterizing an outbreak of catastrophic proportions (see Eq.~\eqref{eq:magT} and discussions therein). 
Nations' intervention efficiency, rated with plus and minus signs, is measured according to the deviation of all countries best linear fit $d=r_0+\log (P^\beta/r_{max}^{dat})$ from the ordinate $r_{max}^{dat}$ of each data point (see the main text for more details).}
\label{tab:mag}
\end{table}
\indent 
Similarly to the Richter scale for seismic events, our epidemic magnitude scale $\mathpzc{T}$ characterizes the {\em local} (since it depends on the population size) strength of an epidemic in an exponential fashion, so that each jump by class identifies a tenfold increase in daily reported infected/deceased among countries with similar population sizes. 
When applied to the available data of the SARS-CoV-2 pandemic, $\mathpzc{T}$ yields the repartition of the epidemic events summarized in Tab.~\ref{tab:mag}, with respect to which we filled the data points in Fig.~\ref{fig:corr} with colors ranging from dark blue to dark red for increasing magnitudes.
%Following the classification above, we filled the data points in Fig.~\ref{fig:corr} with colors ranging from dark blue to dark red for increasing epidemic magnitudes, whose values are listed in Tab.~\ref{tab:mag}. 
A primary observation is that all the countries of our dataset have experienced epidemic events of magnitude equal or larger than $\mathds{III}$~\cite{nota2}, reflecting the severe and broad impact that the SARS-CoV-2 pandemic has had worldwide. 
We notice also that the different responses that countries with similar population sizes (e.g., $P\sim10^7$) have had to the pandemic spread is nicely captured by $\mathpzc{T}$, with e.g.\ cases like Greece or Hungary both experiencing light outbreaks, and cases like Israel or Switzerland facing instead moderate to strong events (see Tab.~\ref{tab:mag}). 
Very strong pandemic events can be instead recognized by the orange colors in countries like Italy, France or Germany, and even more extreme ones by increasingly red colors describing the cases of the United States (magnitude $6.5$) and Spain (magnitude $6.6$). 
Surprisingly, we find that the first-wave epidemic event in Spain features in fact a larger magnitude than the one reported in the United States.\\
\indent 
The linear regression analysis in Fig.~\ref{fig:corr} provides with additional information the classification by magnitude of the SARS-CoV-2 pandemic types.
Unlike other catastrophic events, in fact, epidemic outbreaks can be influenced at their early stages by social intervention strategies~\cite{maier2020effective,ferguson2006strategies,germann2006mitigation,adolph2020pandemic,morse2006next,mate2020evaluating,meidan2020alternating,karin2020adaptive,block2020social,aleta2020evaluation}, and so do their magnitude scales. 
%Social distancing and lockdown protocols~\cite{meidan2020alternating,mate2020evaluating,karin2020adaptive} remain the most prominent and non-pharmaceutical~\cite{morse2006next,aledort2007non} policies to counter the spread of viral agents by slowing their transmission among the population. 
While $\mathpzc{T}$ measures the impact of the pandemic across different countries, the deviation of all countries best linear fit $d=\log(r_{max}/r_{max}^{dat})$ from the ordinate $r_{max}^{dat}$ of each data point can be adopted to rate the effectiveness of the intervention strategies adopted. 
%In particular, by considering the ordinate $r_{max}^{dat}$ of each country and its corresponding value on the linear regression line, we can quantify the relative quality of intervention through the distance $d=r_0+\beta\log P-r_{max}'$. 
Values of $d$ for the available data and reported in Tab.~\ref{tab:mag}, are typically dispersed in the unit interval (dashed green and red lines in Fig.~\ref{fig:corr}) around zero, suggesting the following rating system for the social intervention strategies adopted: $\boldsymbol{+\!+\!+}$) for $d\geq1$ as reported in countries that, despite their population size, managed to contain the outbreak very efficiently; $\boldsymbol{++}$) if $0.5\leq d\leq0.9$ and $\boldsymbol{+}$) if $0\leq d<0.5$ for countries with efficient to good intervention; $\boldsymbol{-}$) if $-0.5\leq d<0$ and $\boldsymbol{--}$) if $-1\leq d<-0.5$ describing instead weak or not prompt responses to the emergent outbreaks. 
We find (see Fig.~\ref{fig:corr}) that highly populated countries like China or Japan applied successful social intervention protocols which kept low the magnitude of the epidemic events, while other countries like Ireland, Belgium or Spain gained less efficient results, experiencing outbreaks of larger magnitudes.\vspace*{+0.1cm}

\begin{figure}[t]
	\includegraphics[width=0.95\linewidth]{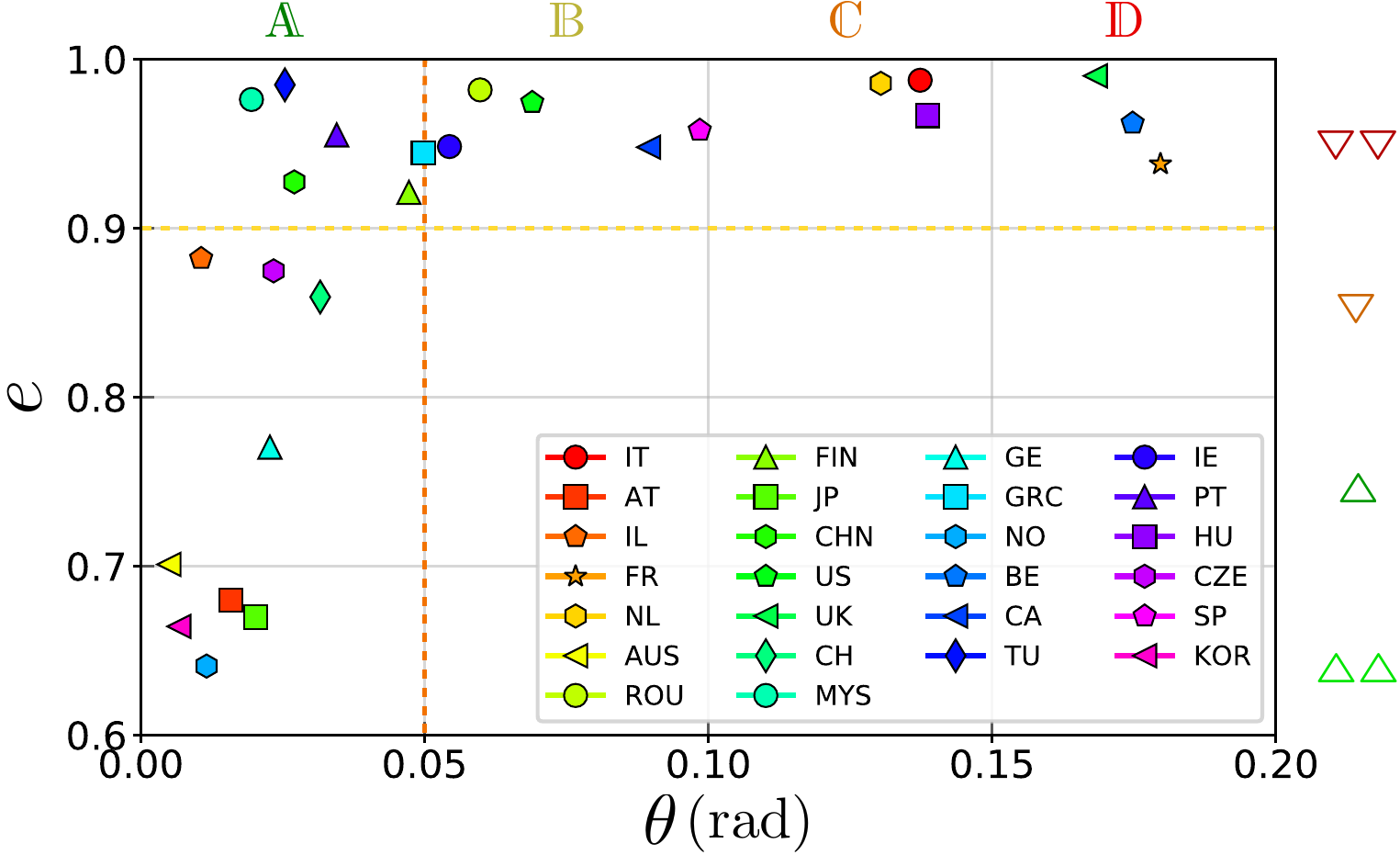}\vspace*{-0.2cm}
	\caption{\textbf{Intensity scales of typical SARS-CoV-2 epidemic trajectories.} (Color online)
	Hybrid scatter plot describing the distribution of typical epidemic events according to their angle $\theta$ (raw data) expressed in radians, and eccentricity $e$ (normalized data).
	%While $\theta$ better catches how critical a country's health system has been, $e$ better rates the strategies adopted for contact tracing of patient identification. 
	Epidemic events with angles larger than the threshold $\theta^*=0.05$ (red dashed line) can be taken to describe {\em fast pandemic} with increasing speed (categories $\mathds{B},\,\mathds{C},\,\mathds{D}$). 
	These categories appear to correspond to epidemic plumes with large eccentricities (class $\text{\smaller[2]$\bigtriangledown\bigtriangledown$}$) and small peak-to-peak separation, reflecting late identification of infected patients or selection of diagnostic testing only for critically ill cases.	
	{\em Slow pandemics} (i.e.\ category $\mathds{A}$), on another hand, disclose a broader spectrum of forms (classes $\text{\smaller[2]$\bigtriangledown\bigtriangledown$},\text{\smaller[2]$\bigtriangledown$},\,\text{\smaller[2]$ \triangle $},\text{\smaller[2]$ \triangle $}\text{\smaller[2]$ \triangle $}$) reflecting different strategies of patient identification.}\vspace*{-0.5cm}
	\label{fig:ecc}	
\end{figure}

\indent
\emph{Outbreak intensity.} 
In the previous section we adopted the epidemic angle $\theta$ (defined by the inclination of $r_{max}$ with respect to the $I$-axes) as an estimator of the ``speed'' of an epidemic event, with large values of $\theta$ reflecting highly critical conditions of the country's health system. 
%Having defined a metric scale for the epidemic magnitude of the SARS-CoV-2 pandemic events, let us now return to the classification of their intensity. 
%We saw that the epidemic angle, $\theta$, obtained by measuring the inclination of the maximal radius $r_{max}$ of the epidemic trajectories, can be adopted to rate the speed of the outbreak based on how critical a country's medical system has been during the spread. 
We have also shown that $\theta$ is linearly correlated (Fig.~\ref{fig:theta_rho}) to the geometric parameter $\rho$, measuring instead the relative width of the epidemic plumes, whose values can be adopted to evaluate the quality of a country's strategies for contact tracing or patient identification. 
The normalization of the trajectories by $r_{max}$ shown in Fig.~\ref{fig:theta_rho}\textbf{a})--\textbf{f}), has highlighted that a further informative quantification of the country's tracking strategies can be obtained by considering the ``eccentricity'', $e=\sqrt{1-\rho^2}$, of the normalized epidemic plumes.  
%In light of its dependence on the eccentricity $e$ of the trajectories via the relation $\rho=\sqrt{2(1-e)}$, the parameter $\rho$ should be a suitable observable for comparing the shapes of the pseudo-ellipses, grouping countries with similar forms and eccentricities. 
%We noticed, nevertheless, that a counter-intuitive effect emerges when performing the selection in classes based on the $(\theta,\rho)$ regions on the normalized trajectories. 
%However, while small (large) values of $\rho$ should describe very narrow (round) pseudo-ellipses, the opposite pattern is observed when {\em normalizing} the trajectories data by their maximal extend, as depicted in Fig.~\ref{fig:theta_rho}\textbf{a})--\textbf{f}). 
%This observation suggests that a more informative comparison of the trajectories forms can be made by measuring the eccentricities of their normalized traces, removing the information about the relative extend of the infected and fatality rates. 
On the one hand, this operation removes all the information about the relative extend of the infected and fatality rates and hence it wipes out the classification of the epidemic events according to the angle $\theta$. 
On the other hand, analyzing the eccentricity of the normalized epidemic plumes yields a more faithful comparison among their forms, clearly isolating narrow plumes from rounder ones. 
To make the best out of this trade-off of information, we have created the hybrid scatter-plot $(\theta,e)$ shown in Fig.~\ref{fig:ecc}, where countries' SARS-CoV-2 events are classified according to the angle of their row epidemic data (just as in Fig.~\ref{fig:theta_rho}) and the eccentricity of their normalized plumes. 
When applied to the available data, this hybrid representation unveils something surprising. 
We find, in fact, that epidemic angles larger than a threshold value $\theta^*\equiv0.05$ always correspond to narrow plumes with normalized eccentricities $e\in(0.9,1)$. 
Instead, plumes with $\theta<\theta^*$ can correspond to a broader spectrum of forms, from the round ones reported in Japan, Austria or Germany, to more narrow types like those of Canada, Portugal or Turkey. \\
\indent 
To explain this surprising pattern, let us notice that the main factor decreasing the eccentricity of a normalized epidemic plume is an increasing time lag between the infected and fatality rate peaks which, in its turn, suggests that the data of reported infected and deceased are less likely correlated. 
The eccentricity can then be adopted for comparing the countries' strategies for patient identification, with small values of $e$ characterizing efficient tracking protocols, and large values of $e$ describing instead situations where the majority of the fatalities were infected patients tested positive only after arriving at the hospitals in critical conditions. 
This is probably best represented in Fig.~\ref{fig:ecc} by cases like the United Kingdom ($e\simeq0.990$), Italy ($e\simeq0.987$) or the Netherlands ($e\simeq0.985$) which all experienced violent outbreaks with identification only of critically ill patients~\cite{onder2020case} and highly critical conditions ($\theta>\theta^*$) of their medical system~\cite{biddison2019too}.\\
\indent 
Fig.~\ref{fig:ecc} suggests therefore the following data-driven classification by {\em intensity}, i.e.\ based on the damage produced on the population, of the SARS-CoV-2 events: 
\begin{itemize}
\item[$\mathds{A}$)] $\theta<\theta^*$, {\em slow} pandemics: low values ($1.4\%$ in Australia to $6.1\%$ in Greece~\cite{repo}) of the largest case-fatality rate, well functioning health system; 
\item[$\mathds{B}$)] $\theta^*\leq\theta<0.1$, {\em moderately}-{\em fast} pandemics: higher ($6.8\%$ in Ireland to $12.2\%$ in Spain) case-fatality rates, mild disruption of the medical system; 
\item[$\mathds{C})$] $0.1\leq \theta<0.15$, {\em fast} pandemics: severe disruption of the health system, possibility of strategic triage, high ($12\%$ in the Netherlands to $14.1\%$ in Hungary) largest case-fatality rate; 
\item[$\mathds{D}$)] $\theta\geq 0.15$, {\em very fast} pandemics: very high ($15.9\%$ in the United Kingdom to $19\%$ in France) largest case-fatality rate, medical crisis, strategic triage. 
\end{itemize}

\indent
Similarly to the magnitude, $\mathpzc{T}$, the epidemic intensity scale can be accompanied by a rating system quantifying in this case the efficiency of patient identification and contact tracing strategies, suitably defined as follows: $\text{\smaller[2]$\triangle\triangle$}$) plumes with eccentricity $e\leq 0.7$, featured by countries that performed extensive testing and rapid patient identification, resulting in less correlated data series for infected and deceased rates as reflected by large peak-to-peak time lags; $\text{\smaller[2]$\triangle$}$) plumes with eccentricities $0.7<e\leq 0.8$, describing good patient identification; $\text{\smaller[2]$ \bigtriangledown $}$) plumes with eccentricity $0.8< e\leq 0.9$, describing mildly efficient patient identification; $\text{\smaller[2]$ \bigtriangledown $}\text{\smaller[2]$ \bigtriangledown $}$) plumes having eccentricity $e\geq 0.9$ and strong peak-to-peak proximity, reflecting weak or not-efficient identification strategies. 
This classification reflects, in a quantitative fashion, a simple yet essential fact regarding the fight against the SARS-CoV-2 virus: rapid patient identification leads to functional health system and low case-fatality rates (category $\mathds{A}^{\text{\smaller[2]$\triangle\triangle$}}$), whereas less and less efficient strategies lead more often to fast pandemics (categories $\mathds{B}$--$\mathds{D}$) depending on the country's hospital capacities.\vspace*{+0.2cm}

%\XXX{all}{ivan}{In principle, we could attach here the discussions, adding a few paragraphs summarizing the theoretical and practical implications that are lengthly discussed in what follows, and close the manuscript.}\\

\par
\emph{4. \underline{Temporal evolution of the epidemic angle.}} 
The introduction of the epidemic magnitude (Fig.~\ref{fig:corr} and Tab.~\ref{tab:mag}) and intensity (Fig.~\eqref{fig:ecc}) enabled a preliminary yet informative classification of the SARS-CoV-2 pandemic events, highlighting the main ingredients underlying their epidemiometric fingerprints. 
An important byproduct of this result lies in the possibility of designing alarm protocols and other precautionary measures to dampen the societal effects of a pandemic event~\cite{rodriguez2007handbook,blaikie2014risk}. 
%In the context of hurricanes, for example, the ever-increasing meteorological techniques combined with a large history of damage created by them over the population, has led governments to design alarm levels~\cite{barber2001tornado,sorensen2000hazard,montz2017natural} to raise the public awareness and prepare towns and cities to better absorb the impact of the storm. 
%In the even more paradigmatic case of earthquakes, knowledge about the magnitude and intensity of the seismic events that have hit a country has led to design warning protocols~\cite{kanamori1997real,aranda1995mexico,kanamori2005real,horiuchi2005automatic} or anti-seismic infrastructures~\cite{reitherman2012earthquakes,achaoui2016seismic} to dampen the earthquakes' impact, lowering its damage on the population. 
In fact, understanding the magnitude and intensity of the SARS-CoV-2 epidemic events could help policy-makers to make informative decisions not only for investing resources to strengthen a country' health system, but it can enhance the public awareness with the design of alarms platforms aiming at facilitating contact tracing or promoting responsible actions of social-distancing. \\
\indent 
Unlike other catastrophic events, in fact, the magnitude and intensity of a pandemic lie entirely in the hands of the countries' governments, their preparedness to absorb the impact of a rapidly emerging outbreak, and the awareness of societies to its potential damage. 
In this light, quick actions at the early stages of an epidemic outbreak in tracking the infected~\cite{moscovitch2020better} or more efficient social intervention protocols~\cite{block2020social,meidan2020alternating} can help curbing down dramatically~\cite{valenti2020social,maier2020effective} its devastating effects. 
This is why an early estimation of the outbreak scales could significantly help designing epidemic alerts for enhancing the public and governmental awareness and help fighting against the virus spreading. 
%Developing warning systems of this kind would require merging the forecasting power of epidemic models~\cite{dehning2020inferring,vespignani2020modelling,cintra2020mathematical} with scales of magnitude and intensity of the pandemic. 
In this context, understanding how the SARS-CoV-2 epidemic scales evolve in time could provide significant information for future design of early warning systems~\cite{kogan2020early} and risk alerts~\cite{pluchino2020novel}.\\
\begin{figure}[t]	
\includegraphics[width=0.95\linewidth]{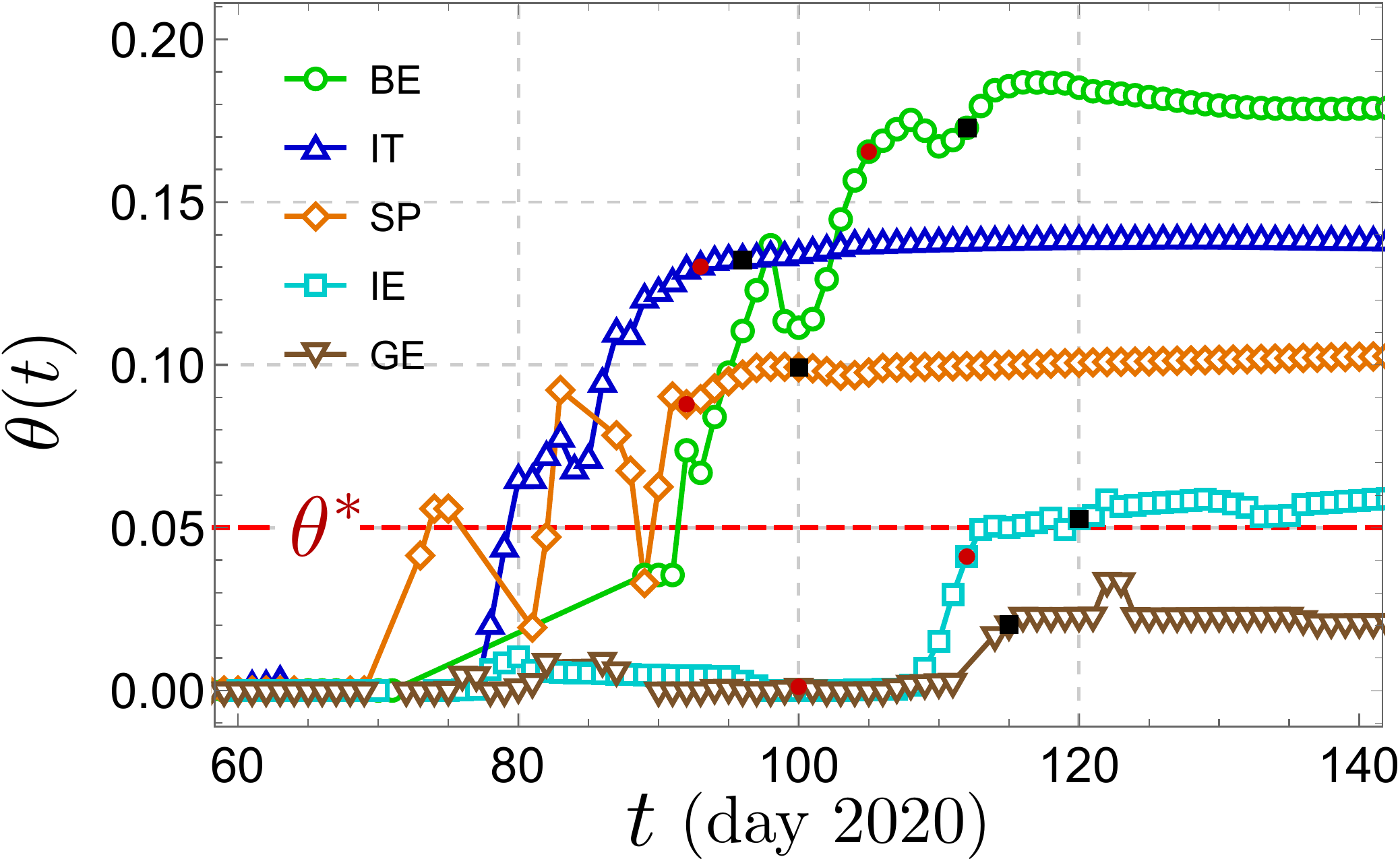}\vspace*{-0.2cm}
	\caption{\textbf{Evolution of the epidemic angle.} (Color online) 
	Temporal evolution of the epidemic angle for different countries' plumes, obtained by best fitting the data series (smoothened by a $15$-day moving average) with a daily update of their epidemic state. 
	Countries experiencing fast epidemic events of category $\mathds{C}^{\boldsymbol{--}}$ or higher cross the threshold angle $\theta^*$ (red dashed line) way before reaching their infected peaks (red filled markers) and fatality peaks (black filled markers). 
	Slow epidemics, like in Germany, keep instead their epidemic angle below $\theta^*$ already when reaching the infected peak.}\vspace*{-0.4cm}
	\label{fig:theta_evo}	
\end{figure}
\indent 
Motivated by this idea, we have analyzed in Fig.~\ref{fig:theta_evo} the temporal evolution of the epidemic angle $\theta$ for a few representative countries of Fig.~\ref{fig:ecc}. 
%There, we noticed that values of the epidemic angle larger than the threshold $\theta^*=0.05$ result into fast outbreaks featuring not-efficient patient identification. 
The results in Fig.~\ref{fig:theta_evo} show that countries having experienced fast pandemics of category $\mathds{C}^{\text{\smaller[2]$ \bigtriangledown $}\text{\smaller[2]$ \bigtriangledown $}}$ or higher such as Italy, Spain or Belgium, all crossed the critical angle $\theta^*$ long before reaching their infected peaks (red filled circles), reflecting a slow responsiveness~\cite{armocida2020italian} to the rapidly emerging outbreak. 
On the contrary, countries like Germany (or Austria and Switzerland, not shown in Fig.~\ref{fig:theta_evo} to simplify the exposition) succeeded in keeping $\theta$ below $\theta^*$ already when they reached their infected rate peaks, suggesting efficient patient identification and well functioning medical systems. 
Advanced knowledge of the epidemic intensity could have helped countries like Italy or Spain in adjusting more rapidly their policies of contact tracing and identification of seriously ill patients, offering more options of intervention to fight the epidemic crisis. 

\begin{figure*}
	\centering
	\includegraphics[width=0.9\linewidth]{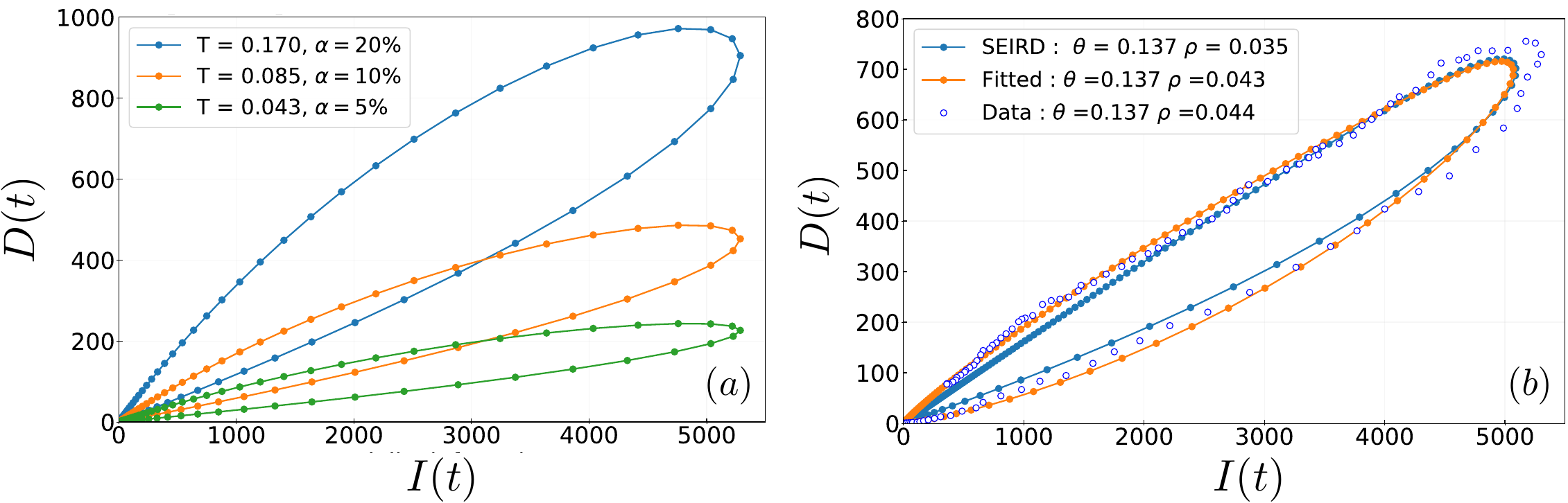}\vspace*{-0.2cm}
	\caption{\textbf{SEIRD trajectories and low-dimensional parametrization.} (Color online) 
	\textbf{a}) Representation in the $(I,D)$ plane of typical (Fig.~\ref{fig:loop_demo}) epidemic trajectories obtained by solving Eqs.~\ref{eq:1} with fixed parameters $(\gamma,\sigma,\mu,R_{0,i},R_{0,f},\kappa,\tau)=(1/3,1/5,1/3,8,0,0,0)$ and decreasing fatality rates $\alpha=(1/5,1/10,1/20)$ (presented as percentages in the legend). This choice of parameters describes an epidemic similar to SARS-CoV-2 but with constant reproduction rate $R_0=4$, where individuals incubate the virus for an average period of $5$ days, gets infected at rate $\beta\simeq 1.33$, and recover/die in an average period of $3$ days. 
	\textbf{b}) Demonstration of the SEIRD trajectory obtained by manual searching for epidemiological parameters matching the data-driven geometric constraints. For the Italian outbreak (open markers), this low-dimensional analysis leads to an SEIRD trajectory (blue curve) characterized by the parameters $(\gamma,\sigma,\alpha,\mu,R_{0,i},R_{0,f},\kappa,\tau)=(1/3,1/5,1/7,1/2.95,3.89,0.98,0.24,44)$ whose features are qualitatively similar to those characterizing the best-fitting pseudo-ellipses (orange curve). }
	\label{fig:seird}	
\end{figure*}

\par
\emph{5. \underline{Low-dimensional parametrization}.} 
%Besides raising key information to design a preliminary metric system for rating the SARS-CoV-2 pandemic types, 
Our geometric analysis additionally hints at an integrated, low-dimensional parametrization for modeling the evolution of real-world epidemic trajectories. 
In fact, the geometric parametrization of the epidemic plumes allows (at least, on the coarse-grained level of the national trends) to curtail with a few parameters the essential information ruling the epidemic spreading~\cite{roda2020difficult,vespignani2020modelling}, out of a variety of epidemiological factors, from e.g.\ different gender-based transmission factors~\cite{wu2020nowcasting,covid2020severe,bi2020epidemiology}, to mobility patterns~\cite{gomez2018critical,soriano2018spreading} and social mixing~\cite{pastor2015epidemic,block2020social}, to cities' pollution~\cite{pluchino2020novel}, different quarantine~\cite{das2020covid} or testing strategies~\cite{hlavacs2020often} and many others. 
Even when adopting a well-mixed approximation among different compartments, developing a predictive framework embracing such a variety of factors easily results into mathematical or computational tour de forces~\cite{arenas2020mathematical,castro2020predictability,dehning2020inferring} whose complexity quickly grows with the number of realistic features included. \\
\indent 
From another viewpoint, modern epidemic models based either on detailed descriptions of the population's compartments~\cite{arenas2020mathematical,maier2020effective,rodriguez2020modelling} or merging simplified versions of the latter with tools of bayesian inference~\cite{dehning2020inferring} and neural networks~\cite{shah2020finding}, typically yield predictions of a country's epidemic trend after fitting {\em either} its data for the infected rates {\em or} the ones fo the fatality rates.
This separated approach leads to best-fitting epidemiological factors that, when adopted to describe the behaviors of other compartments which have not been fitted to the data, lead to results far from the real-world trends. \\
\indent
Our geometric approach offers a viable solution to circumvent both these limitations by best-fitting the synthetic epidemic plumes directly to the data-driven ones, enabling in this way an {\em integrated} and {\em low}-{\em dimensional} estimation of the epidemiological parameters best describing the outbreak evolution. \\
\indent 
As a demonstrative example, let us consider an SEIRD epidemic model~\cite{anderson1992infectious,hethcote2000mathematics} to describe the SARS-CoV-2 dynamics. 
For completeness, let us recall that the SEIRD model suitably characterizes the spreading of a viral agent featuring a latent (sometimes also referred to as cryptic~\cite{davis2020estimating}) phase where susceptible individuals ($S$) become exposed ($E$), i.e.\ they acquire the infection but are not yet infectious. 
After a characteristic incubation period $1/\sigma$ days (with $\sigma\in(0,1]$), exposed individuals become infectious ($I$) and start spreading the disease at a speed, $\lambda$, controlling the average number of people an infected person infects per day. 
Infected individuals spread the disease during an average period of $1/\gamma$ days (with $\gamma\in(0,1]$), after which they either recover ($R$) or die ($D$). 
Let $1/\mu$ (with $\mu\in(0,1]$) be the characteristic period of days during which an infected individual becomes critically ill and eventually dies, and $\alpha\in[0,1]$ the fatality rate characterizing instead the probability of going from infected to death (i.e.,\ $\alpha$), and from infected to recovered (i.e.,\ $1-\alpha$). 
The parameters $\lambda, \sigma, \gamma, \mu, \alpha$ define respectively the infectious, incubation, recovery, mortality and fatality rates of the SEIRD model; to simplify the analysis, let us assume that once recovered, individuals gain immunity. 
The above epidemic process is summarized in the system of differential equations: 
\begin{equation}\label{eq:1}
\begin{aligned}
S'&= -\lambda sI,\quad
E' = \lambda s I - \sigma E \\
I' &= \sigma E -(1-\alpha) \gamma I - \alpha \mu I,\\
R' &= (1-\alpha) \gamma I,\quad
D' = \alpha \mu I,
\end{aligned}
\end{equation}
\noindent
where $s\equiv S/N$ is the density of susceptible individuals and $N$ the population size. 
As a last ingredient, let us include in Eq.~\ref{eq:1} the additional information of a time dependent infectious rate $\lambda(t)$, identifying the introduction of social distancing measures and quarantine strategies. 
These, in fact, aim at lowering the basic reproduction number $R_0=\lambda/\gamma$ of the virus from a certain initial value $R_{0,i}>1$ to a final one $R_{0,f}<1$. 
To model this decay, whose speed will depend on the efficiency of the lockdown strategy applied by a country, let us adopt a logistic function of the form 
\begin{equation}\label{eq:2}
R_0(t)=\frac{R_{0,i}-R_{0,f}}{1+e^{\kappa(t-\tau)}}+R_{0,f},
\end{equation}
\noindent 
where $\tau,\,\kappa\geq0$ are two ``intervention'' parameters ruling respectively the time of the inflection point in the profile of $R_0(t)$ (i.e.\ the day of the country's main lockdown) and the decay rate of the reproduction number as a result of the lockdown efficacy. 
In particular, values of $\kappa\sim\mathcal{O}(1)$ result into a fast decay of $R_0(t)$ (signaling a rapid and efficient intervention), while $\kappa\sim10^{-1}$ or smaller results into a very slow convergence towards $R_{0,f}$.\\
\indent 
Already at this simplistic level, the epidemiological $(\lambda,\sigma,\gamma,\mu,\alpha,R_{0,i},R_{0,f})$ and intervention $(\tau,\kappa)$ parameters aiming at representing the data-driven epidemic plumes, identify an $8$-dimensional phase space for the dynamical system in Eq.~\eqref{eq:1}. 
Thanks to our geometric parametrization, it proves possible to find the functional dependences relating this variety of parameters to the three geometric factors $(r_{max}, \theta, e)$, whose relations identify a series of data-driven parametric constraints to reduce the degrees of freedom of the problem. 
Finding the exact dependence between the epidemiological variables and the geometric factors is beyond the scope of the present work and will be discussed elsewhere. 
Nevertheless, it is immediate to verify that e.g.\ an increase of the case-fatality rate $\alpha$ leads to an increase of the epidemic angle (Fig.~\ref{fig:seird}\textbf{a})---i.e.\ larger epidemic intensities---while increasing the infection rate $\lambda$ yields larger values of $r_{max}$---i.e.\ larger epidemic magnitudes---supporting our heuristic arguments in Sec.~3. \\
\indent 
%By combining these functional relations with the values of $(r_{max},\theta,e)$ obtained from the data-driven trajectories, one can impose parametric constraints identifying a sub-manifold of the $8$-dimensional phase space where the evolution of the dynamical system in Eq.~\eqref{eq:1} lies. 
Fig.~\ref{fig:seird}\textbf{b} demonstrates this integrated, low-dimensional approach applied to the Italian epidemic data. 
We have performed a manual search of the epidemiological parameters best fitting the data-driven values $\theta_{IT}\simeq0.137$ and $\rho_{IT}\simeq0.44$, selecting a suitable sub-manifold of the phase space featuring trajectories geometrically congruent to the data-driven one. 
%This reduces by two the dimension of the dynamical system's phase space, which could be further reduced by considering additional information coming e.g.\ from the different curvilinear velocities of the right and left lobes of the trajectory, or the local curvature of the trajectory
Even if they do not best-fit the data-driven plumes (empty markers and blue curve in Fig.~\ref{fig:seird}\textbf{b}), the set of parameters we identified generates a SEIRD plume nicely matching the epidemic data. 
In particular, we find that a relatively large initial reproduction number $R_{0,i}^{IT}\simeq3.9$ (in qualitative agreement to the more precise ones obtained by best fitting regional data~\cite{d2020assessment}) and a low lockdown efficiency parameter $\kappa_{IT}\simeq0.24$, reflecting the slow decay already observed in Fig.~\eqref{fig:skewness} and therein quantified by a high infected rate skewness. 
The low-dimensional analysis further discloses another realistic ingredient characterizing the Italian pandemic type, i.e.\ the high case-fatality rate $\sigma_{IT}\simeq14.2\%$ reflecting the strong intensity of the Italian outbreak and the critical conditions reached by its health care system. 
Further developing these geometric-based concepts could lead to the identification of additional parametric constraints further reducing the degrees of freedom of epidemic models, possibly boosting their forecasting power. 
Considering additional information coming e.g.\ from the different curvilinear velocities of the right and left lobes of the epidemic plumes, the local curvature of their traces may yield unfamiliar perspectives in achieving this task. 

\par
\emph{6. \underline{Future directions}}. 
Our study is only a preliminary step in the design of metric systems for epidemic events. 
We expect that the results will inspire the development of more refined epidemiometric frameworks for rating the magnitude and the impact of present and future epidemics, helping governments and other decision-makers to strengthen their policies of containment and better respond to such extreme events. 
In this perspective, we highlight in what follows a few important directions of future research in which respect we believe that our geometric approach could be further developed. 

\begin{figure*}
	\centering
	\includegraphics[width=0.42\linewidth]{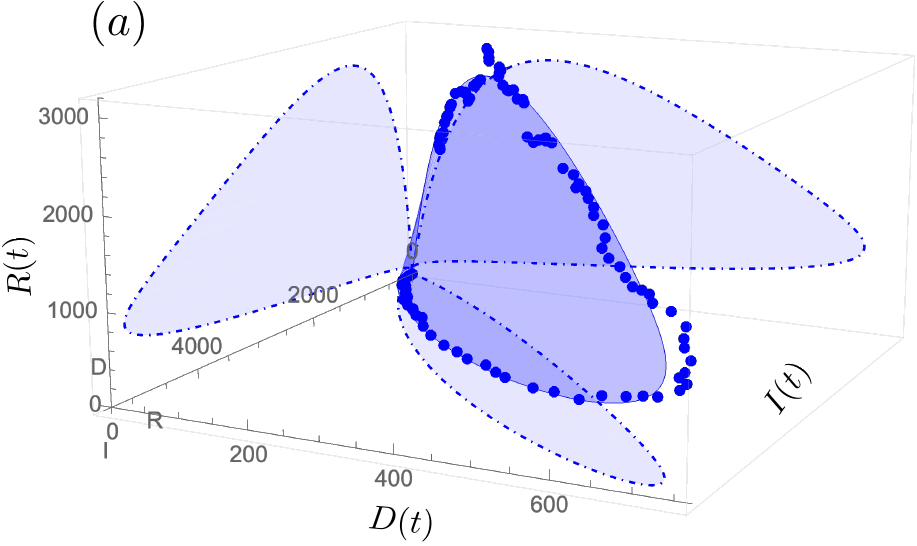}
	\includegraphics[width=0.42\linewidth]{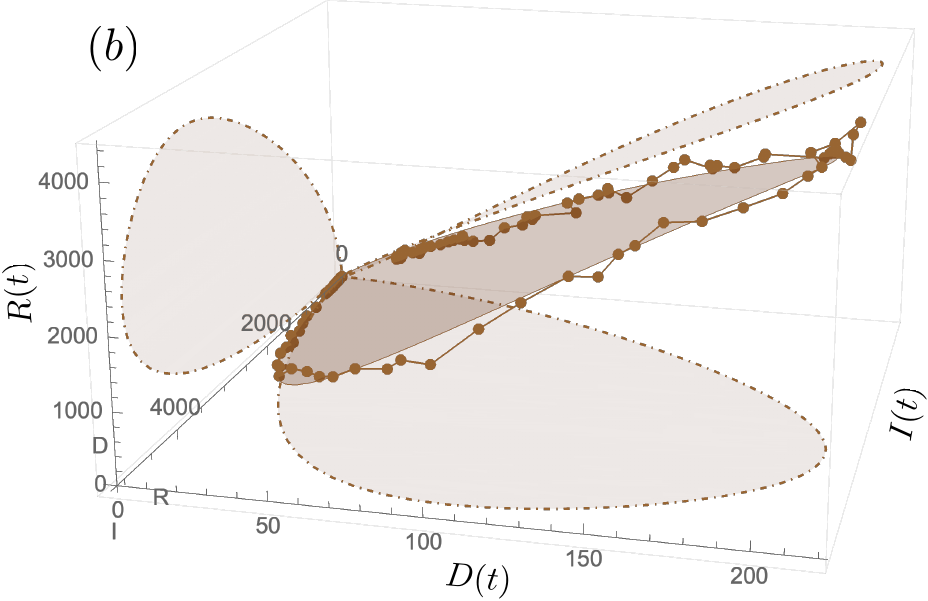}\\
	\includegraphics[width=0.42\linewidth]{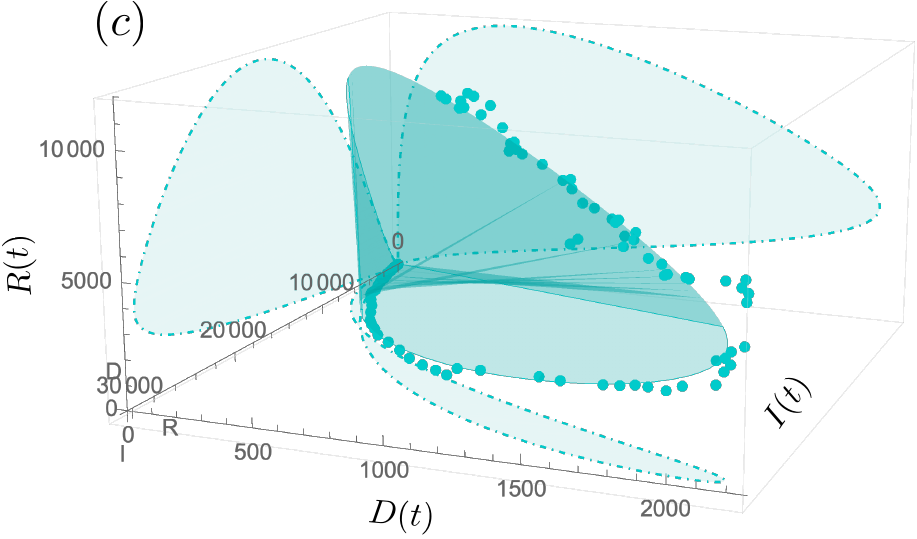}
	\includegraphics[width=0.42\linewidth]{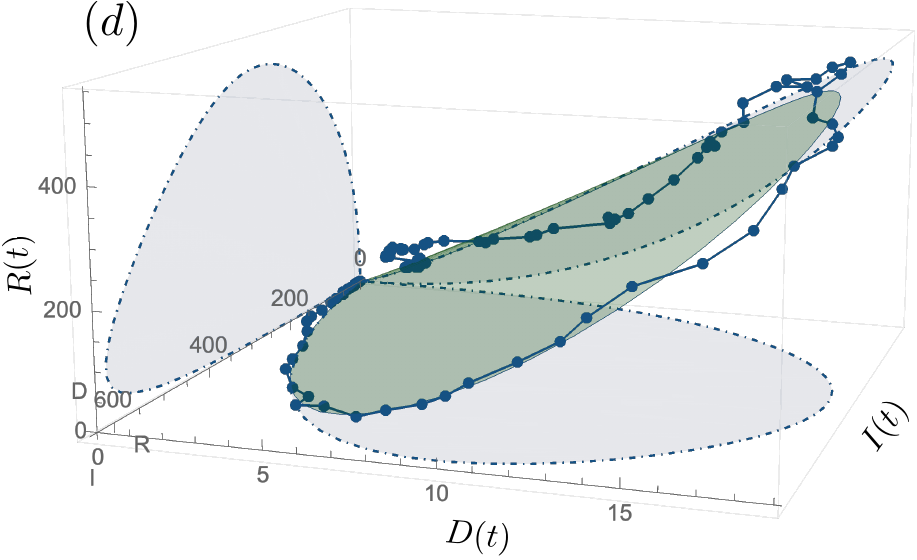}\\
	\caption{\textbf{SARS-CoV-2 epidemic surfaces.} 
	Evolution in the $(I,D,R)$ space of the data (markers) describing the epidemic state for a selection of countries: \textbf{a}) Italy, \textbf{b}) Germany, \textbf{c}) United states and \textbf{d}) Austria. 
	Data have been selected to identify the first-wave event for each country, where a second one was observed.
	Each trajectory traces a surface (shadowed areas) whose projections on the $(I,D)$, $(I,R)$ and $(R,D)$ planes are further shown. 
	While Italy's surface unfolds in the space almost perpendicular to the $(I,D)$ plane (where it traces the narrow trajectory discussed in Fig.~\ref{fig:seird}\textbf{b}), Germany's surface unfolds nearly parallel to the $(I,D)$ plane, tracing instead a rounder pseudo-ellipses (i.e.\ larger normalized eccentricity). 
	Behaviors respectively similar to \textbf{a}) Italy and \textbf{b}) Germany are observed in \textbf{c}) the United States, as well as in Spain or France (with the last two not shown in the above to ease the exposition) and in \textbf{d}) Austria, as well as Switzerland or Australia (not shown). 
	}\label{fig:3D}	
\end{figure*}

\par
\emph{i}) \emph{Beyond first-waves}. 
Our analysis focused on typical (i.e.\ counter-clockwise evolving) epidemic trajectories describing the first-wave events of the SARS-CoV-2 pandemic in a selection of countries. 
As of July 20 2020, some nations included in our analysis have entered secondary epidemic waves whose magnitude appears already to be larger than the classification in Tab.~\ref{tab:mag}; this is the case of e.g.\ the United States, Iran, Israel or Serbia. 
Extending the geometric method to secondary waves is possible as long as the first and the second-wave peaks of the infected and deceased rates are sufficiently resolved over time (Fig.~\ref{fig:secwav}\textbf{a}). 
In this case, fitting the data by multiple (skewed) Gaussians allows to trace a new epidemic trajectory in the $(I,D)$ plane whose evolution may cross itself and disclose geometric features significantly different from those of the first wave. 
By isolating the fitting functions describing each lobe (Fig.~\ref{fig:secwav}\textbf{a}), it is possible to perform an analysis perfectly analogous to the one described above for the case of first-waves, identifying the epidemic angle $\theta$, largest extend $r_{max}$ and eccentricity $e$ of the second-wave trajectory. 
As most of the nations that entered the second wave have not yet reached their new infected peak, we cannot yet draw conclusive evaluation about the magnitude and intensity of their new epidemic events, but we track the evolution of the geometric parameters characterizing their new trajectories as suggested in Sec.~5 (see Fig.~\ref{fig:theta_evo}). 
In Fig.~\ref{fig:secwav}\textbf{b} we have presented the epidemic trajectories describing the state of Israel. 
A preliminary evaluation shows that while the new wave has already a magnitude $\mathpzc{T}_2\simeq6.6$, i.e.\ a very strong epidemic event comparable to the first-wave event of Spain or US (see Sec.~3), its epidemic angle is still relatively low ($\theta_{IL}<0.005$) with respect to the threshold value $\theta^*=0.05$ of the onset of a fast pandemic (Fig.~\ref{fig:ecc}). 
Comparing the magnitude and intensity of different epidemic events in the same country could help understanding how different countries prepared themselves to absorb and dampen the impact of a new wave of epidemic events.

\par
\emph{ii}) \emph{Including daily testing data}. 
An important direction of future research for improving the epidemic measures defined in the above concerns the inclusion of the information related to the number of daily testing performed by each country. 
Depending on their resources, in fact, different countries applied different strategies of testing or contact tracing in their fight~\cite{cohen2020countries} against the spread of SARS-CoV-2. 
Germany or Russia, for example, have been performing a tremendous amount of tests since the early stages of the pandemics and over a very broad~\cite{stafford2020covid} fraction of the population, including both symptomatic and asymptomatic patients. 
Countries like Italy~\cite{onder2020case} or Spain~\cite{tanne2020covid} had instead to prioritize their diagnostic capacities to test patients with more severe clinical symptoms and in need of hospitalization. 
Including this relevant bit of information in the analysis above, e.g.\ by normalizing the number of new daily infected by the corresponding number of daily new tests, would possibly result into an even more meaningful comparison of the countries epidemic magnitude than the one in Tab.~\ref{tab:mag} and, likewise, of the classification by intensity depicted in Fig.~\ref{fig:ecc}. 

\par
\emph{iii}) \emph{Zooming-in: local characterization of countries pandemic types.} 
Our results have focused on analyzing the epidemic trends reported at national levels, offering a country-to-country comparison of their SARS-CoV-2 pandemic types. 
The epidemiometric system of epidemic events proposed however, can be equivalently adopted to analyze more local datasets of each country, offering a magnifying lens to determine the magnitude and the intensity of the epidemic events observed within states, regions or even on the smaller scales of provinces and towns. 
In the case of Italy, for example, regions such as Lombardia, Emilia-Romagna, Piemonte and Veneto have suffered more severe epidemic events than the rest of the country~\cite{pluchino2020novel}, and a similar situation has been reported in United States for the states of New York, California and (more recently) of Florida~\cite{jernigan2020update}. 
An efficient intervention at the national level could find its crucial ingredients in a rapid intervention on the level of its states, regions or provinces. 
Combined with forecasting tools and prior epidemic risk assessments, our epidemiometric framework could provide the design of local epidemic alerts to enhance the awareness of governments and inhabitants already at local levels, helping to counter the spreading of highly infectious viruses like SARS-CoV-2. 

\par
\emph{iv}) \emph{Geometry of the epidemic surfaces}.
As a conclusive remark, let us notice that in our developed geometric analysis, we have focused only on the projections of the epidemic trajectories in the $(I,D)$ plane. 
However, other compartments (e.g.\ recovered, critical patients, active cases, etc) can be included in the analysis, resulting in a multi-dimensional representation of the epidemic state of each country. 
For instance, the addition of the compartment of daily recovered ($R$) yields the emergence of new information to further refine the classification of the pandemic fingerprints reported in different nations. 
Fig.~\ref{fig:3D}) contains a few examples of ``epidemic surfaces'' in the $(I,D,R)$ 3D-space observed in a selection of countries. 
In particular, the cases of Italy (Fig.~\ref{fig:3D}\textbf{a}) and Germany (Fig.~\ref{fig:3D}\textbf{b}) clearly exhibit striking differences in their dynamic evolution: Italy's epidemic surface features a narrow cross-section in its projection on the $(I,D)$ plane and a broad trajectory in the $(R,D)$ plane, as opposed to the epidemic surface characterizing Germany's outbreak. 
The two surfaces, in fact, appear to be roughly orthogonal with each other, a clear indication that in Italy the increase of infected yielded a rapid and simultaneous increase of deceased. 
Similar behaviors can be found in cases belonging to the same magnitude and intensity. 
For example, Austria (Fig.~\ref{fig:3D}\textbf{d}) as well as Norway or Switzerland shares similar patterns to those observed in Germany, while countries like USA, France or Spain feature inclinations of their epidemic surfaces resembling the one observed in Italy. \\
\indent 
The identification of the patterns shared by the outbreaks of different countries would have otherwise been impossible if we had to limit our view to the classical time evolution of the epidemic compartments as in Fig.~\ref{fig:skewness}. 
Exploring epidemic dynamics from this novel, geometric-based perspective could unveil new ``hidden'' features characterizing their evolution and foster new methods for their statistical analysis and mathematical modeling. 
We expect that our geometric framework and results will inspire alternative approaches to the study of epidemic evolution, possibly leading to longer-termed forecasting techniques or to more refined epidemiometric systems for the design of epidemic alerts and early warning systems. 

\par
\emph{7. \underline{Discussion}}. 
We have presented a geometric framework to analyze and systematically classify the trajectories of the SARS-CoV-2 pandemic across different countries in the $(I,D)$ plane via three geometric parameters $(r_{max},\theta,e)$. 
Our geometric measures enables the design of a preliminary epidemiometric system to quantify the magnitude of a country's outbreak and its intensity, resembling respectively the Richter and the Mercalli measures for seismic events, and further adding information about the efficacy of lockdown strategies and of patient identification. 
The epidemic scale measures we defined help identifying a spectrum of SARS-CoV-2 pandemic types, ranging from weak epidemic events with slow speed, like those reported in Japan, Australia or South Korea (magnitude $\mathpzc{T}\simeq3.4$, class ${\boldsymbol{++}}$, category $\mathds{A}^{++}$), to very extreme events with intense damage inflicted on the population, like the cases of United Kingdom (magnitude $\mathpzc{T}\simeq5.9$, class ${\boldsymbol{--}}$, category $\mathds{D}^{\boldsymbol{--}}$) or Italy (magnitude $\mathpzc{T}\simeq6.0$, class ${\boldsymbol{-}}$, category $\mathds{C}^{\boldsymbol{-}}$). \\
\indent 
However, unlike other catastrophic events, the magnitude and intensity of an epidemic event entirely depends on the responsiveness of the countries' government and the capacity of their medical systems, jointly with the awareness of their population to their potential damage. 
In this respect, early estimation of the epidemic scales (e.g.\ by merging them with forecasting models) could significantly contribute to the design of warning systems~\cite{pluchino2020novel,kogan2020early} or protocols for virus alerts to enhance the public and governmental responsiveness. 
We showed that, in cases like Italy or Spain (Fig.~\ref{fig:theta_evo}), the epidemic angle $\theta$ has crossed  the threshold $\theta^*=0.05\mathrm{rad}$ from slow to fast pandemics way before reaching the infected and fatality peaks, reflecting a slow responsiveness to the rapidly emerging crisis. 
From the mathematical perspective, our geometric method further raises relevant insights to improve current epidemic models. 
The geometric characterization of the epidemic plumes in the $(I,D)$ plane, discloses an integrated and low-dimensional approach to modeling the epidemic trends by imposing geometric constraints relating the data-driven parameters $(r_{max},\theta,e)$ to the epidemiological factors entering the epidemic models adopted. 
This allows to lower the number of independent parameters, identifying a suitable sub-manifold of the high dimensional phase space where plumes congruent to the data-driven ones can be found. 
We have demonstrated this approach by manually searching, in an SEIRD model with a time-dependent reproductive number, the best choices of the epidemiological and intervention parameters satisfying the data-driven geometric constraints for the Italian trends, obtaining a realistic description of the reported behaviors. \\
\indent 
We foresee that the merging of our data-driven, low-dimensional approach with more advanced mathematical or computational methods~\cite{dehning2020inferring,shah2020finding}, could lead to predictions of the epidemiometric fingerprints of real-world epidemics with ever-increasing accuracy, possibly disclosing new directions to the identification of optimal priors for more efficient and longer-termed epidemic forecasting.

\par
\emph{\underline{Acknowledgements}}
S.H.\ thanks the Israel Science Foundation, ONR, the BIU Center for Research in Applied Cryptography and Cyber Security, NSF-BSF Grant no.\ 2019740, and DTRA Grant no.\ HDTRA-1-19-1-0016 for financial support. I.B.\ thanks A.\ Collini and M.\ Hidalgo Soria for valuable discussions. 

\bibliographystyle{unsrt}
\bibliography{mybib}

\begin{thebibliography}{10}
\bibitem{nota1}
By definition $e:= \sqrt{1-(b/a)^2}$, with $a$ and $b$ respectively the major
  and minor radiuses of a regular ellipse. in our case $r_{max}$ and
  $r_{\perp}$ play respectively the analogous roles of $a$ and $b$ for the
  plume-like epidemic trajectories, from which the expression in the main text
  follows.
\bibitem{nota2}
Notice that the linear regression analysis yields an intercept $r_0\simeq-4.1$
  which, in turns, means a proportionality factor $\mathcal{A}_0\simeq10^{-4}$
  between $r_{max}$ and $p$, meaning a magnitude of $\mathpzc{T}\simeq3.2$ for
  $p\sim10^5$, and increasing to $\mathpzc{T}\simeq6.3$ for $p\sim10^8$.
\end{thebibliography}

\end{document}